\documentclass[12pt,a4paper]{article}
\usepackage{epsfig}
\usepackage{mathtools}
\usepackage{amsfonts}
\usepackage{amssymb}
\usepackage{amsmath}
\usepackage{color}
\usepackage{slashed}
\usepackage{cite}
\usepackage{caption}
\usepackage{subcaption}

\usepackage{geometry} 
\usepackage{graphicx}
\graphicspath{{pictures/}}

\usepackage{multirow}           
\usepackage{array}              

\def\la   {\langle}
\def\ra   {\rangle}

\def\Li2  {\text{Li}_2}
\def\LL   {\mathcal L}
\def\ep {\varepsilon}
\def\im {\text{Im}}
\def\tr {\text{tr}}
\def\bk {\mathbf{k}}
\def\erf {\mbox{erf}}
\def\unren {\text{unrenom}} 
\def\MSbar {\overline{MS}} 
\def\gE {\gamma_E}

\begin{document}

\begin{center}

\vspace{3cm}

{\bf \Large NNLO  corrections to false vacuum decay rate 
	in thin-wall approximation } \vspace{1cm}

{\large M.A. Bezuglov$^{1,2}$ and A.I. Onishchenko$^{1,2,3}$}\vspace{0.5cm}

{\it $^1$Bogoliubov Laboratory of Theoretical Physics, Joint
	Institute for Nuclear Research, Dubna, Russia, \\
	$^2$Moscow Institute of Physics and Technology (State University), Dolgoprudny, Russia, \\
	$^3$Skobeltsyn Institute of Nuclear Physics, Moscow State University, Moscow, Russia}\vspace{1cm}
\end{center}

\begin{abstract}
Recently it was discovered that the Standard Model vacuum may suffer an essential metastability near the Planck scale. In this regard it makes sense to study in more detail the decay processes of the metastable vacuum and to develop methods for their more accurate analysis. In this article we develop a technique to calculate two loop radiative corrections to false vacuum decay rate in thin-wall approximation and apply it to false vacuum decay in scalar quantum field theory with cubic and quartic interactions. The results obtained use dimensional regularization and given in two different renormalization schemes: Coleman-Weinberg and $\MSbar$. 
\end{abstract}

\begin{center}
\end{center}

\newpage

\tableofcontents{}\vspace{0.5cm}

\renewcommand{\theequation}{\thesection.\arabic{equation}}

\section{Introduction}

In modern studies of vacuum physics, an important role is played by the phenomenon of quantum tunneling of vacuum from one state to another. The most studied process is the decay of metastable vacuum otherwise known as false vacuum decay.  From the thermodynamical point of view this decay is a first-order phase transition. The description of such transitions in the framework of quantum field theory first appeared in \cite{Coleman1,Coleman2,Coleman3,Kobzarev}. The latter proceed through the nucleation of true vacuum bubbles inside a false vacuum. If the size of these bubbles is greater than some critical value they begin rapidly expand, so that over the time the whole universe is in a state of true vacuum. 

The most interesting example of vacuum decay, which attracted a lot of attention, is a possible metastability of the electroweak vacuum in the SM  (at large scale around $\sim 10^{11}$ GeV) \cite{SMmetastability1,SMmetastability2,SMmetastability3,SMmetastability4,SMmetastability5,SMmetastability6,SMmetastability7,SMmetastability8}. The recent studies are not limited by SM alone but also include its extensions, see \cite{SMtunneling1,SMtunneling2,SMtunneling3,SMtunneling4,SMtunneling5,SMtunneling6,SMtunneling7,SMtunneling8,SMtunneling9,
SMtunneling10,SMtunneling11,SMtunneling12,SMtunneling13,SMtunneling14}
and references therein. There is also an extravagant possibility that the vacuum in the universe is not in a lowest energy state but nevertheless is completely stable. The possibility of such scenario was considered in \cite{Gonzalez:2017hih}. Moreover, first-order phase transitions have probably occurred in early universe. For this, there are strong theoretical foundations in the form of Sakharov conditions for realization of observed baryon-anti-baryon asymmetry \cite{Sakharov}, see also \cite{Morrissey}, that require in addition to violations of CP and baryon number symmetries the breakdown of thermal and chemical equilibrium at some time in the universe. And the first order phase transition is just a good option for this. Traces of such processes taking place in the early universe may be seen even now in the form of stochastic gravitational waves \cite{Witten,Kosowsky1,Kosowsky2,Kamionkowski,Caprini} and as such could be detected in the future gravitational-wave experiments \cite{Cai}.
 
The false vacuum decay is realized by the appearance of an imaginary part in its energy. The latter in its turn could be expressed through the imaginary part of the partition function. There are various methods, which may be used to calculate it, such as potential deformation method \cite{KleinertPathIntegrals,IntroQuantumMechanics,QFTCriticalPhenomena,InstantonsLargeN,WeinbergClassicalSolutions,PrecisionDecayRates} and direct computation via path integrals \cite{DirectApproachQuantumTunneling,PrecisionDecayRates}. At the tree level, the decay rate is given by the exponent of the action ($\sim e^{-S_b}$) calculated for a single extremal trajectory solving classical Euclidean equations of motion. This type of extremal trajectory is called bounce solution. Higher order radiative corrections are then calculated as perturbative expansion around this solution. The one loop contribution to decay rate could be easily expressed through the ratio of two functional determinants corresponding to the Gaussian fluctuations around bounce solution and false vacuum. To evaluate functional determinants we may choose from  Green function methods \cite{KleinertChervyakov1,KleinertChervyakov2,GreenFunctionScalarQFT,GarbrechtMillington2,GarbrechtMillington3,GarbrechtMillington4,GarbrechtMillington5}, heat kernel methods \cite{heatkernel-usermanual,CalculationsExternalFields,MassiveContributionsQCDtunneling,DunneFunctionalDeterminants}, or use famous Gel'fand-Yaglom theorem \cite{GenfaldYaglom} and methods based on it generalization \cite{KirstenMcKane1,KirstenMcKane2}.
The two loop radiative corrections to the false vacuum decay in for dimensional quantum scalar field theory were first calculated in \cite{VacuumDecaySFT} using generalization of technique first developed for quantum mechanics by\cite{Instantons2loop,Olejnik,Instantons3loop,Instantons3loop-SineGordon,QuantumThermalFluctuationsQM,VacuumDecayQM}. Beyond one loop the corresponding techniques were also developed in the investigation of effective Euler-Heisenberg Lagrangian \cite{EulerHeisenberg1,EulerHeisenberg2,EulerHeisenberg3} and computation of quantum corrections to classical string solutions \cite{Tseytlin1,Tseytlin2}.

This article has two goals. First, it contains a more detailed description of the method for calculating two loop radiative corrections to false vacuum decay in thin wall approximation \cite{VacuumDecaySFT} in $\MSbar$ renormalization scheme. Secondly, we give also results for one and two loop radiation corrections in  Coleman-Weinberg (CW) renormalization scheme. The section \ref{VacuumDecay} and its subsections contain all the main material of the present paper. First, we describe our model with single scalar field experiencing cubic and quartic self-interactions  and false vacuum decay in it at one-loop order. Next, in subsection \ref{GreenFunctionsubsection} we present required expressions for  Green functions in bounce background and at false vacuum in a planar thin wall approximation. All of these results are already known from \cite{GreenFunctionScalarQFT} and \cite{VacuumDecaySFT}. In subsection \ref{renormalization} we introduce details of  the renormalization schemes employed. Then, in subsections \ref{OneLoopDecay} and \ref{TwoLoopDecay}  we present the details of one and two loop calculations in dimensional regularization and in two different renormalization schemes: Coleman-Weinberg and $\MSbar$. Finally in section \ref{Conclusion} we come with our conclusion. The appendices \ref{Effective potential} and \ref{Wave function renormalization} contain technical details about the calculation of renormalization counterterms at false vacuum, while appendix \ref{sunset-appendix} contains the details of calculation of the most complicated two-loop sunset diagram. 

\section{False vacuum decay in scalar field theory}\label{VacuumDecay} 

The most convenient way to describe false vacuum decay rate is through the imaginary part of ground state energy 
\cite{Coleman1,Coleman2,Coleman3} :
\begin{equation}
\Gamma = -2 \im E_{FV}/\hbar
\end{equation}
The latter could be easily obtained with the use of partition function 
\begin{equation} 
Z = \mbox{tr} e^{-T H/\hbar}
\end{equation}
as
\begin{equation} 
E_{FV}= -\lim\limits_{T\rightarrow\infty}\frac{\hbar}{T}\log Z .
\end{equation}
Considering imaginary part of partition function as small compared to its real part we then get
\begin{equation} 
\Gamma =  -2 \im E_{FV}/\hbar = \lim\limits_{T\rightarrow\infty} \frac{1}{T}\left|\frac{\mbox{Im} Z}{\mbox{Re} Z}\right| .
\end{equation}
In the case of four dimensional scalar field theory the required partition function could be written in terms of Feynman path integral as ($Z = Z[0]$):
\begin{equation}
Z [J] = \int [d\Phi] \exp \left[
-\frac{1}{\hbar} \left(
S [\Phi] - \int d^4 x J(x) \Phi (x)
\right)
\right] ,
\end{equation}
where the Euclidean action $S[\Phi]$ for scalar field is given by
\begin{equation}
S[\Phi] = \int_{-T/2}^{T/2} d^4 x \left[
\frac{1}{2} (\partial_{\mu}\Phi)^2 + U(\Phi) .
\right]
\end{equation}
Evaluating imaginary part\footnote{Real part of partition function is evaluated by saddle point at false vacuum} of partition function in one bounce and saddle point approximation\footnote{See  \cite{Coleman2,SMtunneling3} for more details}, that is if we take into account only Gaussian fluctuations, we get  \cite{Coleman2,SMtunneling3} ($\tau$ is Euclidean time):  
\begin{equation}
\label{decay_unrenormalezed}  
\frac{\Gamma}{V} = 
\left(\frac{S}{2\pi\hbar}\right)^2\left|\frac{\mbox{det}'\left(-\frac{d^2}{d\tau^2}-\Delta+U''(\varphi)\right)}{\mbox{det}\left(-\frac{d^2}{d\tau^2}-\Delta+U''(v)\right)}\right|^{-\frac{1}{2}}\exp\left(-S/\hbar\right)\Big\{1+\mathcal{O}(\hbar)\Big\}
\end{equation}
Here $\varphi$ denotes bounce solution, $v$ - false vacuum solution,  $\mbox{det}'$  means that zero modes were removed from the determinant and factor $\left(\frac{S}{2\pi\hbar}\right)^2$ is obtained by integrating over zero modes with the use of collective coordinates of the bounce. This result is valid only at one-loop approximation and to go further we should go beyond Gaussian fluctuations.
Note, that here and below, the parameter $\hbar$ is introduced simply in order to count loops.

\begin{figure}[h]                         
\begin{minipage}[h]{0.47\linewidth}         
\center{\includegraphics[width=1.0\linewidth]{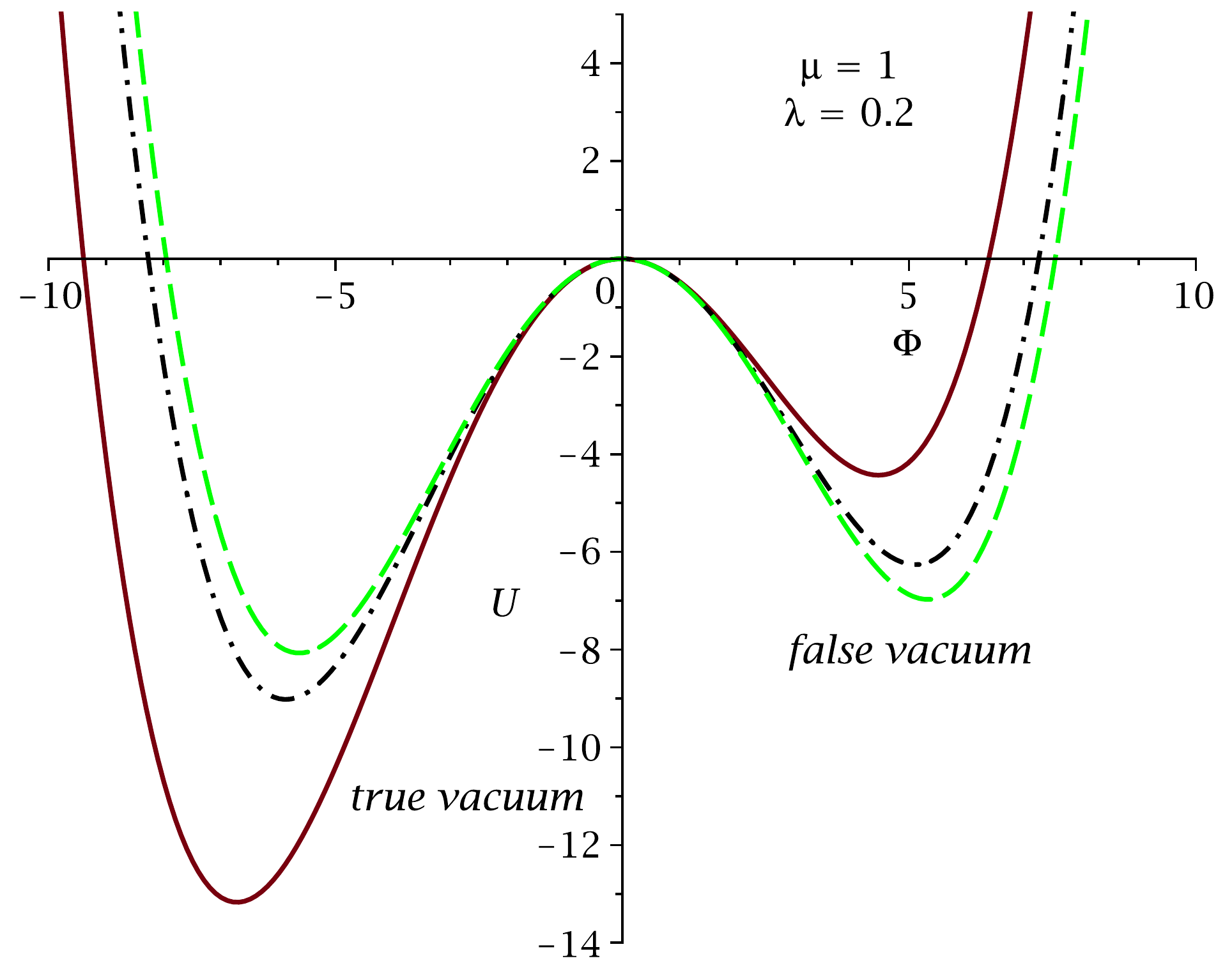}} 
\end{minipage}
\hfill                                      
\begin{minipage}[h]{0.47\linewidth}
\center{\includegraphics[width=1.0\linewidth]{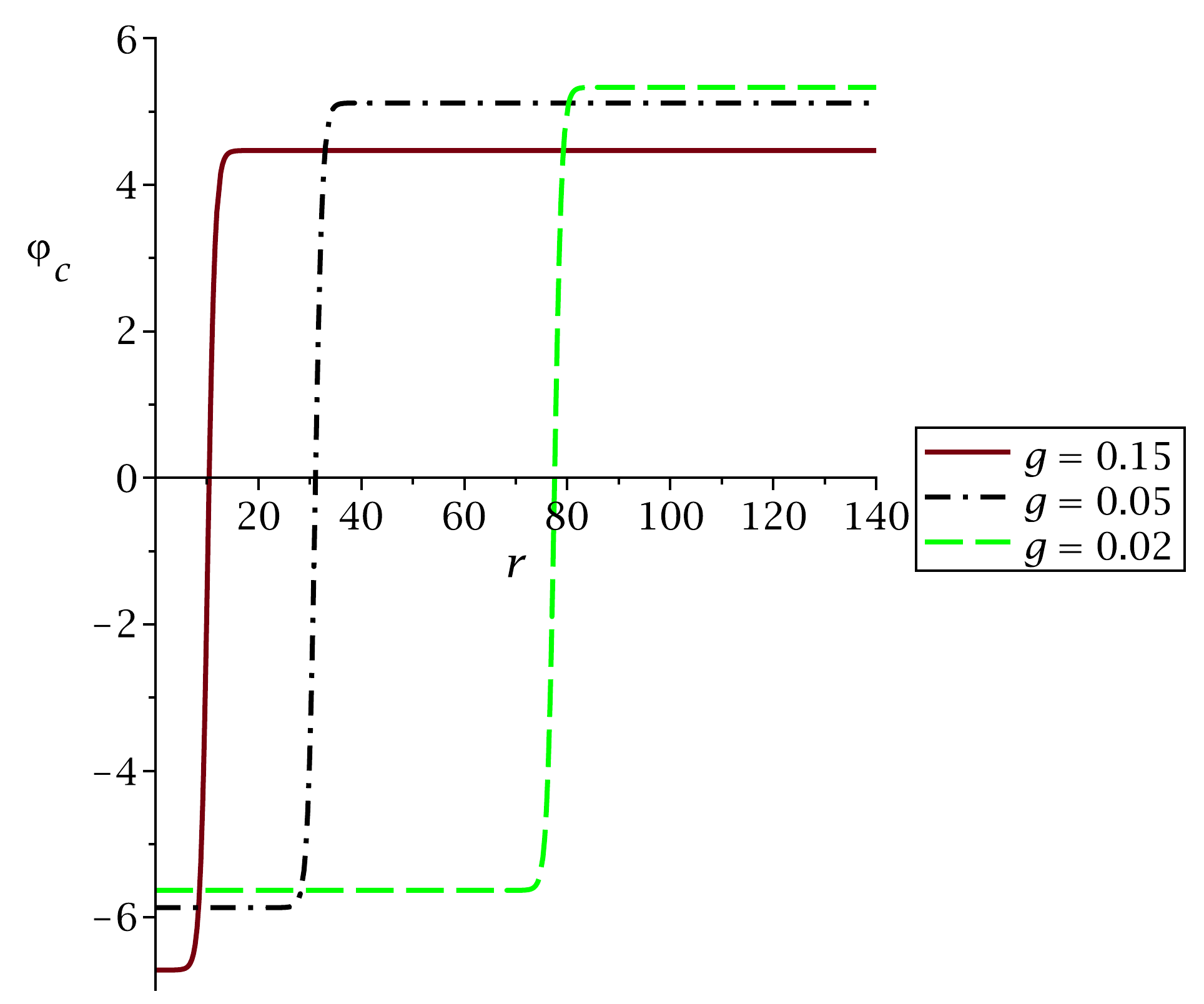}}
\end{minipage}
\caption{Potential $U(\Phi)$ on the left and the corresponding bounce solution on the right, Numerical solutions for equation \eqref{4DBounceEq} were obtained using the program AnyBubble\cite{AnyBubble}} \label{PotQFT}  
\end{figure}

In this work we will consider the false vacuum decay for scalar field theory with cubic and quartic interactions of the form
\begin{equation}
U(\Phi) = -\frac{1}{2}\mu^2 \Phi^2 + \frac{g}{3!}\Phi^3 + \frac{\lambda}{4!}\Phi^4 + U_0
\end{equation}
The potential $U$ has two non-degenerate minima $\varphi = v_{\pm}$ with
\begin{equation}
v_{+}=-\frac{3g}{2\lambda}-\sqrt{v^2+\frac{9g^2}{4\lambda^2}}, \qquad v_{-}=-\frac{3g}{2\lambda}+\sqrt{v^2+\frac{9g^2}{4\lambda^2}}
\end{equation}
and difference in potential levels:
\begin{equation}
U(v_{-})-U(v_{+})=\sqrt{3}g\left(\frac{v^2}{3}+\frac{3g^2}{4\lambda^2}\right)^{\frac{3}{2}}=\frac{v^3g}{3}+\mathcal{O}(g^2)
\end{equation}
In the limit $g\to 0$ the minima at $v_{\pm} = \pm v$ become degenerate. In what follows it will be convenient to chose $U_0 = \frac{\mu^2 v^2}{4} - \frac{g v^3}{6}$, so that the potential vanishes in the false vacuum $\varphi = +v$.
The plots of the potential $U(\Phi)$ for different values of $g$ are shown in the left part of Fig. \ref{PotQFT}. In the scalar theory under consideration the bounce  corresponds to $O(4)$-symmetric solution of the classical equation of motion
\begin{equation}
-\partial^2\varphi + U' (\varphi) = 0\, , \label{eqmotion}
\end{equation}
where $'$ denotes the derivative with respect to field $\varphi$. It is a four dimensional bubble of radius $R$ separating true vacuum inside it from a false vacuum outside. To solve equation (\ref{eqmotion}) we first rewrite it in hyperspherical coordinates  
\begin{equation}
-\frac{d^2}{dr^2}\varphi - \frac{3}{r}\frac{d}{dr}\varphi + U' (\varphi)\, . \label{4DBounceEq}
\end{equation}
Then the numerical solution of this ordinary differential equation could be obtained for example with the use of shooting method \cite{AnyBubble} and the right part of Fig. \ref{PotQFT} shows bounce solutions for several values of the coupling $g$. 

The part $\frac{3}{r}\frac{\partial\varphi}{\partial r}$ in equation (\ref{4DBounceEq}), in analogy with mechanics, corresponds to friction. Because of this part equation (\ref{4DBounceEq}) can not be solved analytically in a general case. Analytical solution exist only for a few number of specific potentials\cite{QFTBounsesAnalytical1,QFTBounsesAnalytical2,QFTBounsesAnalytical3,QFTBounsesAnalytical4}. Nevertheless, some analytic approximation can still be done in the case when the true and false vacuums are practically degenerate, so that we can neglect the cubic term in the potential.   This limit is known as the thing wall approximation and was considered for example in \cite{GreenFunctionScalarQFT}. Thing wall approximation also allows us to neglect the dumping term in the equation of motion, so that it could be easily solved and  the bounce is given by the well-known kink solution \cite{NonperturbativeMethodsReview}:
\begin{equation}
\label{KinkSolution}
\varphi (r) = v \tanh [\gamma (r-R)]\, , \quad \gamma = \frac{\mu}{\sqrt{2}}\, , \quad v= 2\gamma\sqrt{\frac{3}{\lambda}}
\end{equation}
The radius $R$ of the bubble is then obtained by extremizing the bounce action \cite{GreenFunctionScalarQFT}:
\begin{equation}
R = \frac{12\gamma}{g v} = \frac{2\sqrt{3\lambda}}{g}\, .
\end{equation}
and we see that in the thin-wall approximation the radius $R$ is very large $R\gg 1$.
The classical action corresponding to found bounce solution can be calculated straightforwardly and is given by
\begin{equation}
\label{BounceAction}
S_{b} = \int d^4 x \left[
\frac{1}{2}\left(\frac{d\varphi}{d r}\right)^2 + U(\varphi) 
\right] = \frac{8\pi^2 R^3\gamma^3}{\lambda}\, .
\end{equation}
 
\begin{figure}[h]                         
\begin{minipage}[h]{0.47\linewidth}         
\center{\includegraphics[width=1.0\linewidth]{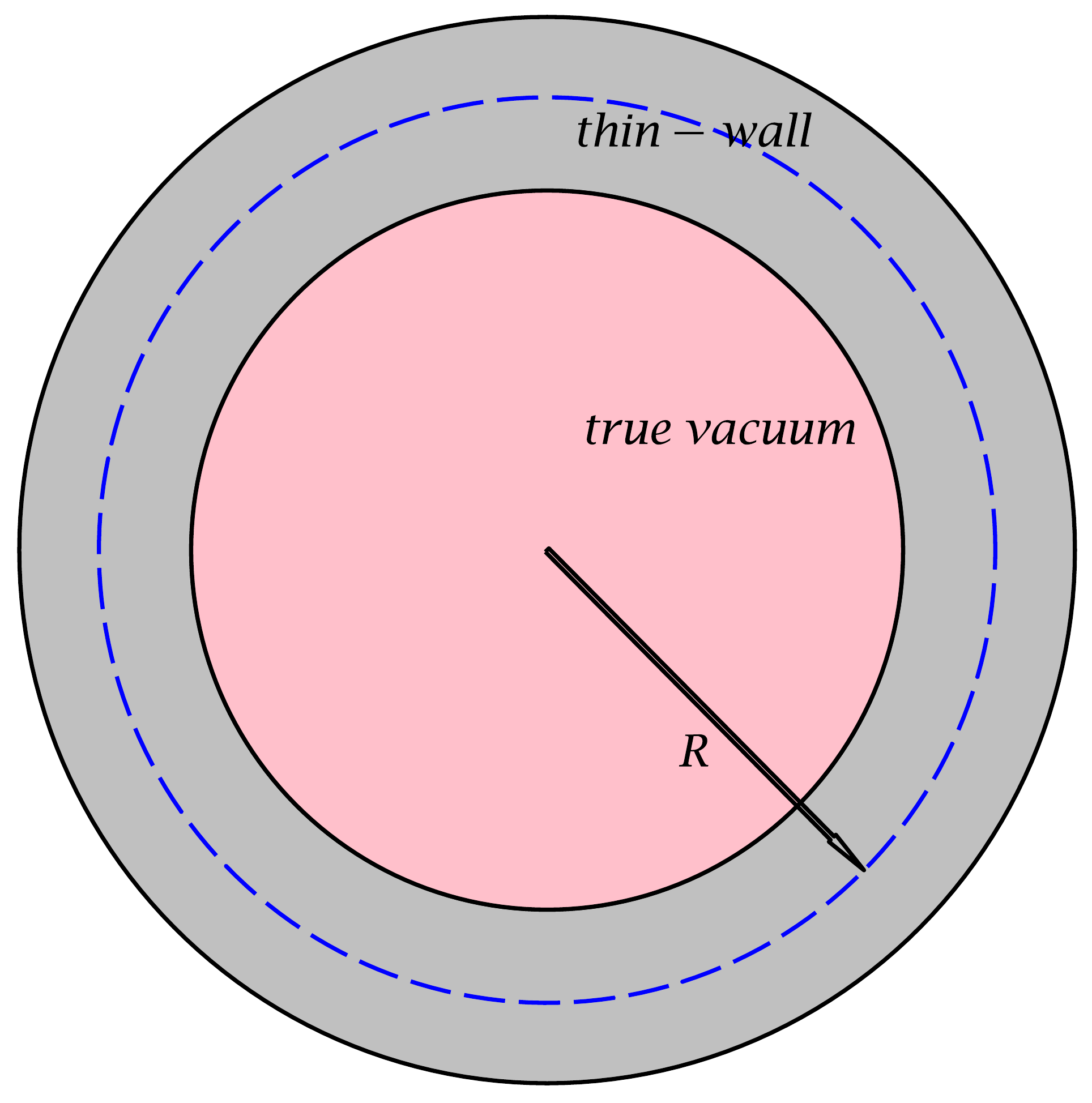}} 
\end{minipage}
\hfill                                      
\begin{minipage}[h]{0.47\linewidth}
\center{\includegraphics[width=1.0\linewidth]{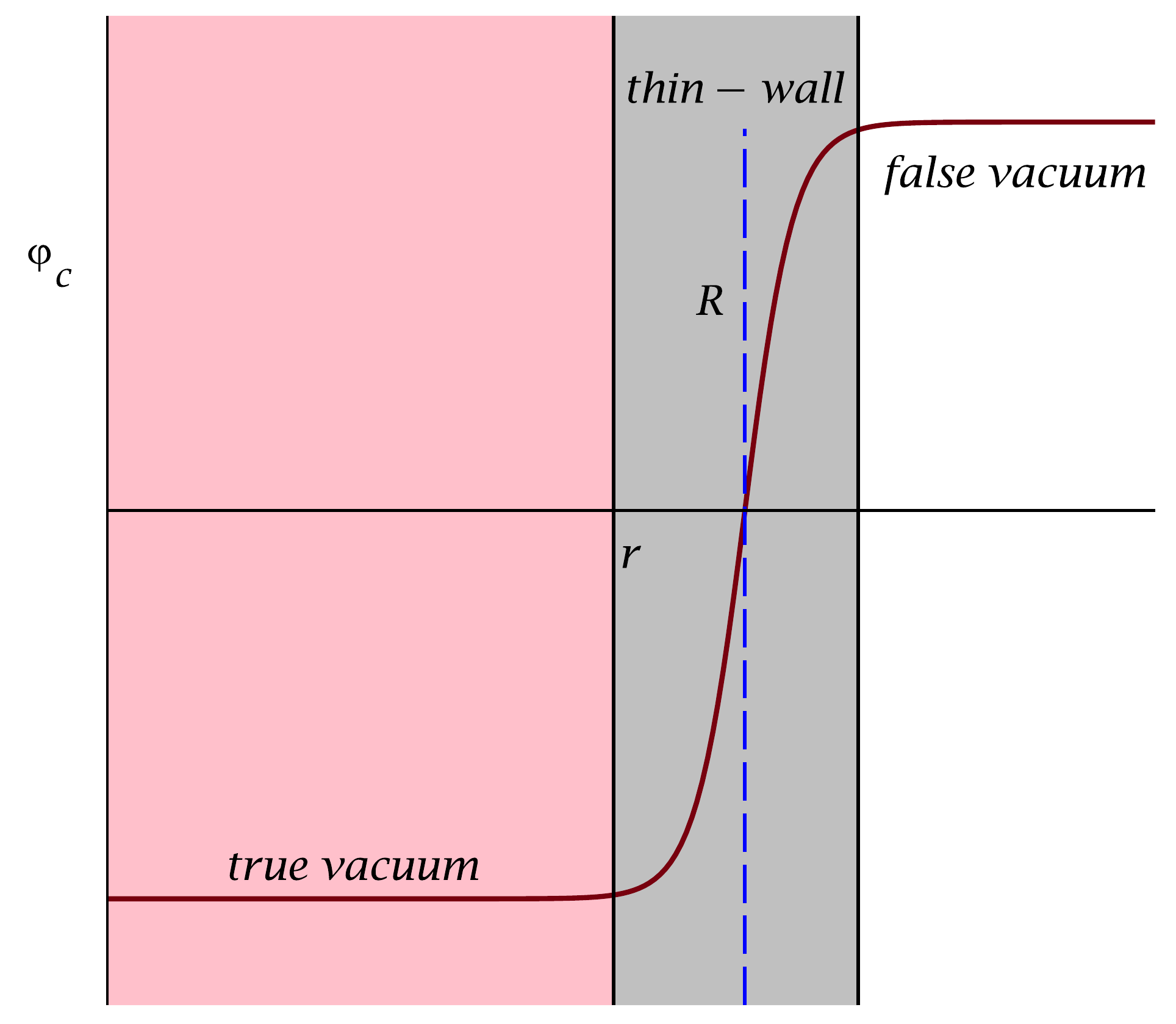}}
\end{minipage}
\caption{General structure of thing wall solution} \label{ThingWallPlot}  
\end{figure}

The structure of the kink solution (\ref{KinkSolution}) is shown schematically in Fig. \ref{ThingWallPlot}. As was already mentioned, this solution describes the bubble of true vacuum separated from the false vacuum by a wall. The thickness of this wall is much smaller then the radius $R$ of the bubble, that is why the limit $g \to 0$ is called the thing-wall or planar-wall approximation.  

\begin{figure}[h]
	\center{\includegraphics[width=0.47\textwidth]{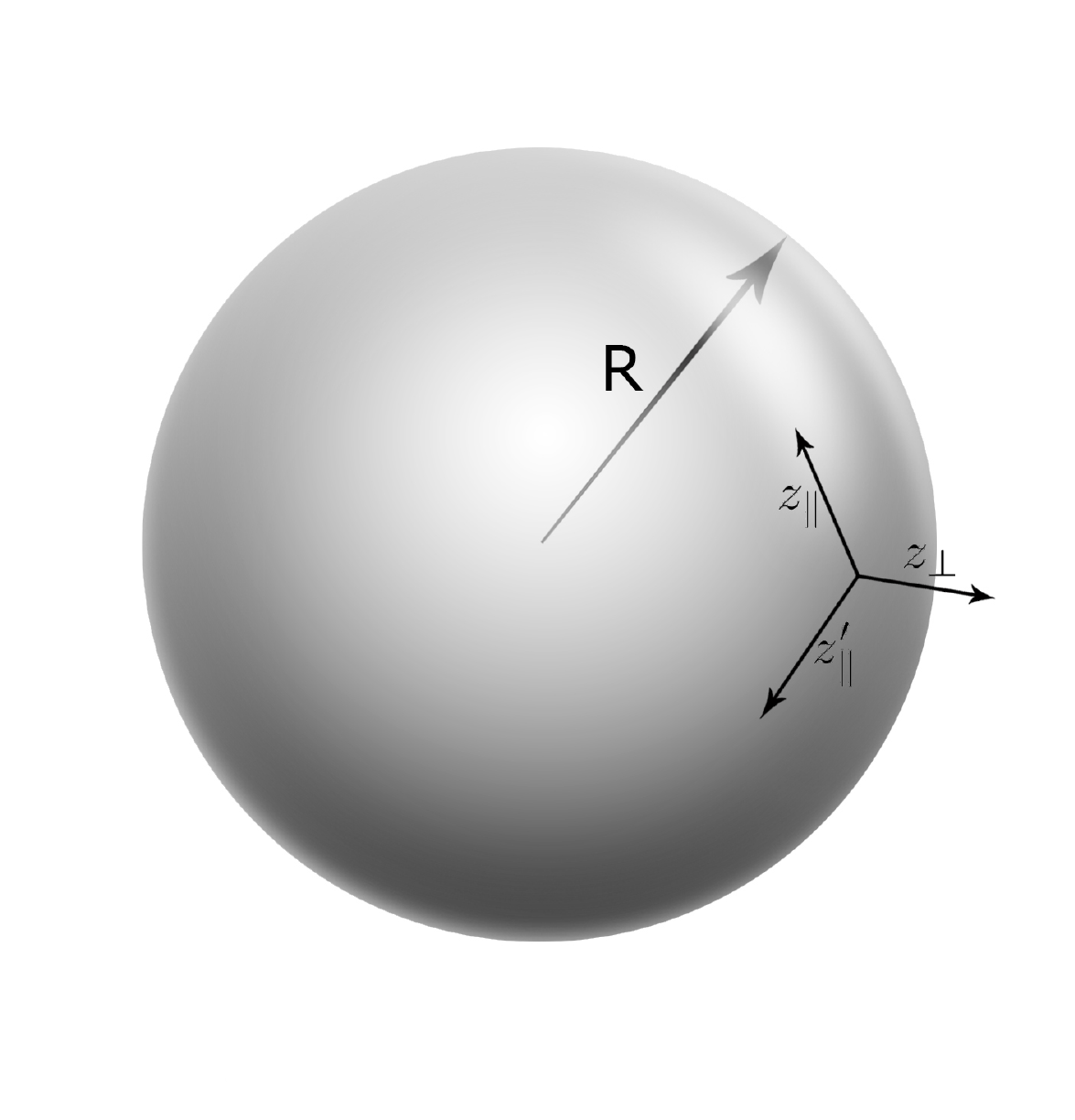}}
	\caption{Coordinate system in planar-wall approximation.}
	\label{planar-wall_coordinates}
\end{figure}

In this approximation it is useful to use coordinate system associated with the bubble wall instead of its center. The corresponding planar-wall coordinate system is shown in Fig. \ref{planar-wall_coordinates},   
where $ \mathbf{z_{\|}}$\footnote{In this paper bold symbols like $\mathbf{a}$ will denote three ($d-1$ in dimensional regularization) dimensional vectors.} is the set of coordinates on $(d-1)$\footnote{As usual in dimensional regularization we have $d=4-2\epsilon$ } dimensional sphere and  $z_{\bot}$ is the coordinate in the direction orthogonal to the latter. In general, space $ \bf{z_{\|}}$ is described by the metric induced on the surface of the sphere, but in our approximation $R$ is very large and  therefore we can assume that the space $ \bf{z_{\|}}$ is Euclidean. The only difference, which should be taken into account, is the finite volume of $ \bf{z_{\|}}$ space, which is the volume of $(d-1)$ - dimensional sphere. It will be convenient to express the latter through the classical action corresponding to the bounce solution $S_b$ (\ref{BounceAction}):
\begin{equation} 
\label{volume}
\int d^{(d-1)}\mathbf{z_{\|}}=\frac{2\pi^\frac{d}{2}}{\Gamma\left(\frac{d}{2}\right)}R^{d-1}=\frac{\lambda}{4\gamma^3}S_b
\end{equation}

We also would like to note that thin wall approximation allows us greatly simplify the integration because the largest values of many integrals will be collected on the above wall, the most common case is the integral
\begin{equation}
\int\limits_0^{\infty}r^3 f(\tanh(\gamma(r-R)))dr=R^3\int\limits_{-1}^{1}\frac{f(x)dx}{\gamma(1-x^2)} + \mathcal{O}(R^2)\, ,
\end{equation}
where $f(z)$ is regular function. Numerical calculations show that this integration strategy gives very good approximation for large values of $R$, already at $R>20$ more than 1\% accuracy is achieved. 

Finally, as to calculate 2-loop corrections to false vacuum decay we will use Green function method it will be convenient to use alternative to \eqref{decay_unrenormalezed}  one-loop expression for decay rate obtained in \cite{GreenFunctionScalarQFT}:
\begin{equation}
\frac{\Gamma}{V} = \left(\frac{S_b}{2\pi\hbar}\right)^2 \frac{(2\gamma)^5 R}{\sqrt{3}} \exp \left[
-\frac{1}{\hbar} S_b + I^{(1)}
\right]\, , \label{decay1loop}
\end{equation}
where
\begin{equation}
I^{(1)} = -\frac{1}{2}\tr^{(5)} \left(\ln G^{-1}(\varphi) - \ln G^{-1} (v)  \right)\, , \label{funcdet1loop}
\end{equation}
$\tr^{(5)}$ denotes trace over only positive-definite eigenmodes, while $G(\varphi)$ and $G(v)$ stand for Green functions in bounce background and in false vacuum correspondingly. The inverse of Green functions $G^{-1}$ at bounce $\varphi$ and false vacuum $v$ are defined as  ($\triangle^{(4)}$ is the four-dimensional Laplacian):
\begin{equation}
G^{-1}(\varphi)\equiv \frac{\delta^2 S[\Phi]}{\delta\Phi^2 (x)}\Bigg|_{\Phi = \varphi} = -\triangle^{(4)} + U'' (\varphi)\, ,
\end{equation}
The spectrum of operator $G^{-1}(\varphi)$ in bounce background
\begin{equation}
(-\triangle^{(4)} + U'' (\varphi))\phi_{n j} = \lambda_{n j} \phi_{n j}
\end{equation}
is given by \cite{GreenFunctionScalarQFT}:
\begin{equation}
\lambda_{nj} = \gamma^2 (4-n^2) + \frac{j(j+2)-3}{R^2} .
\end{equation}
We see, that the latter contains one negative mode at $\lambda_0 = \lambda_{20}$ and four zero modes at $\lambda_{21}$. The "continuum" of positive-definite modes starts at $\lambda_{10}\approx \lambda_{11} = 3\gamma^2$. It is precisely the presence of negative mode in bounce background, which is responsible for false vacuum decay.

\subsection{Green function in bounce background}\label{GreenFunctionsubsection}

\begin{figure}[h]                         
\begin{minipage}[h]{0.47\linewidth}         
\center{\includegraphics[width=1.0\linewidth]{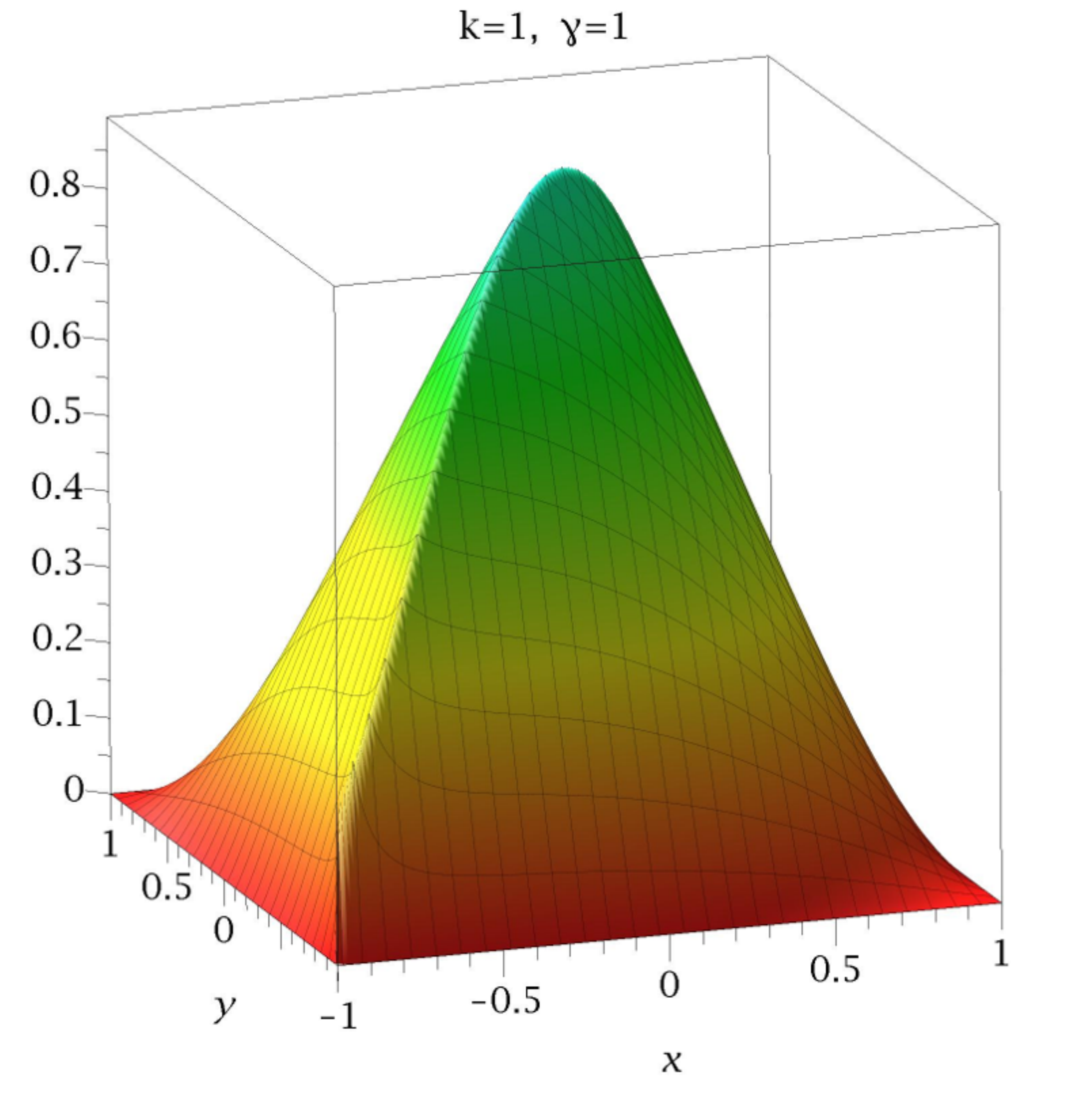}} 
\end{minipage}
\hfill                                      
\begin{minipage}[h]{0.47\linewidth}
\center{\includegraphics[width=1.0\linewidth]{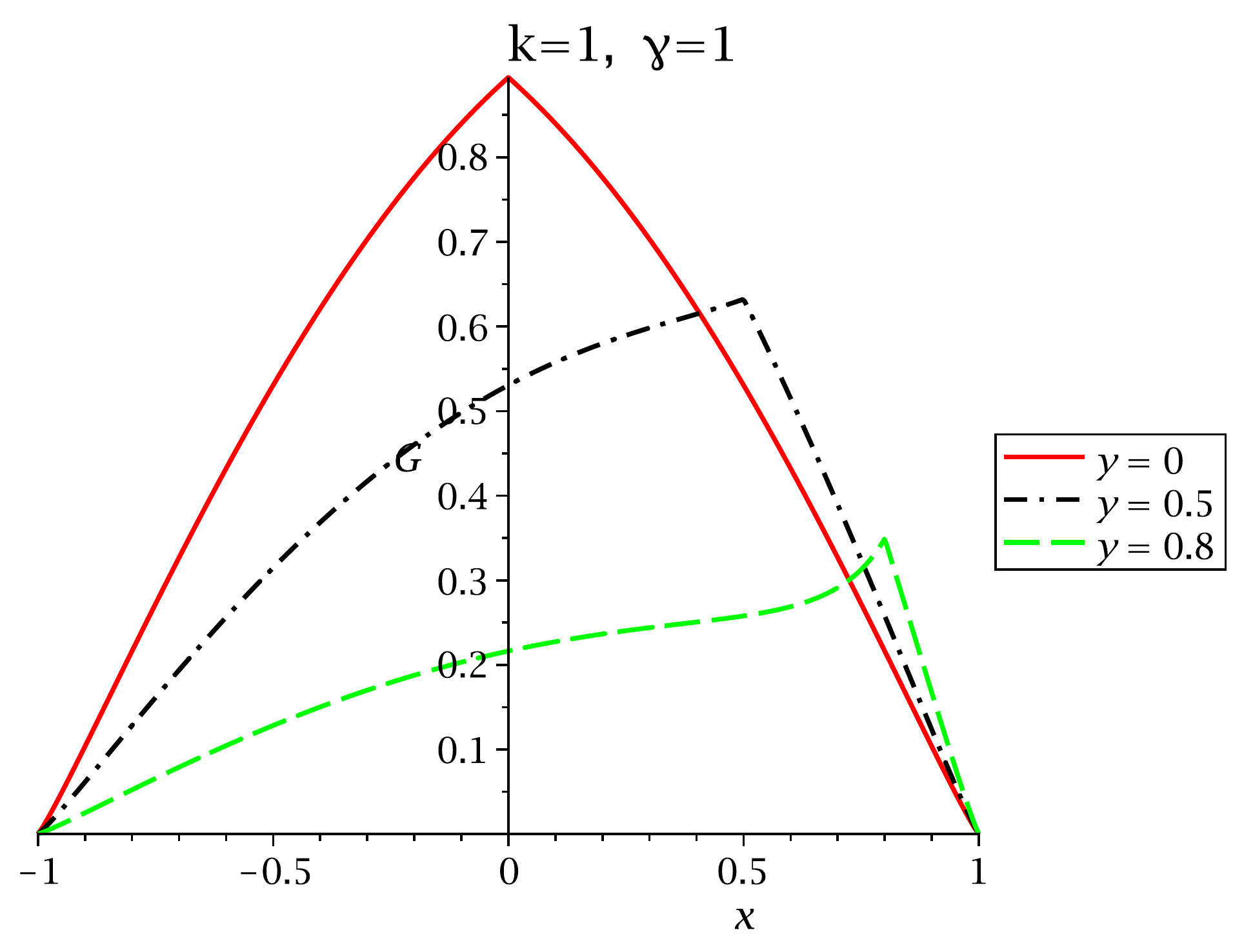}}
\end{minipage}
\caption{Plots of the Green function in the case $m>2$, here $k$ stands for $|\mathbf{k}|$. } 
\label{GreenGen1}  
\end{figure}

\begin{figure}[h]                                   
\begin{minipage}[h]{0.47\linewidth}
\center{\includegraphics[width=1.0\linewidth]{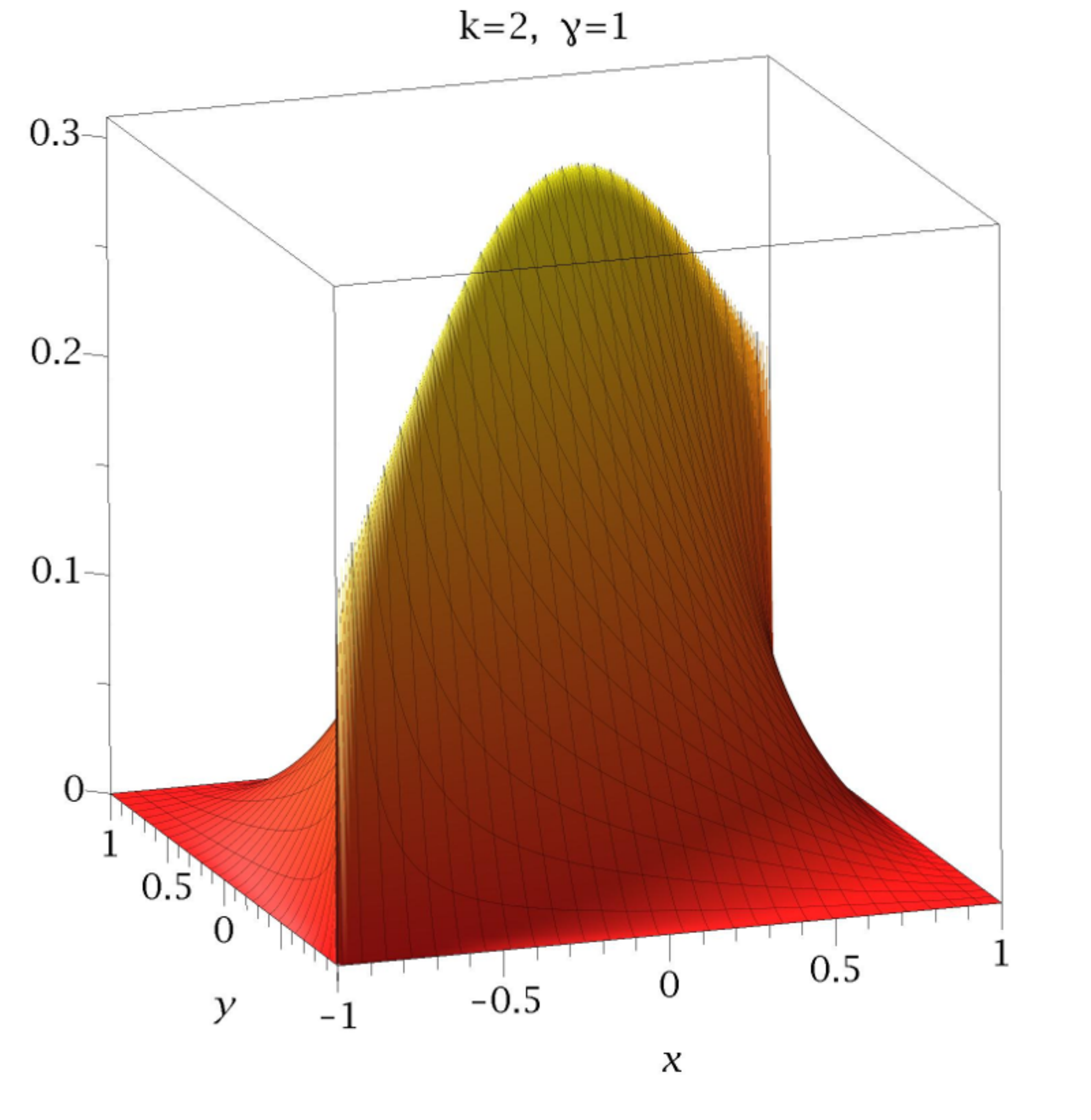}}
\end{minipage}
\hfill
\begin{minipage}[h]{0.47\linewidth}
\center{\includegraphics[width=1.0\linewidth]{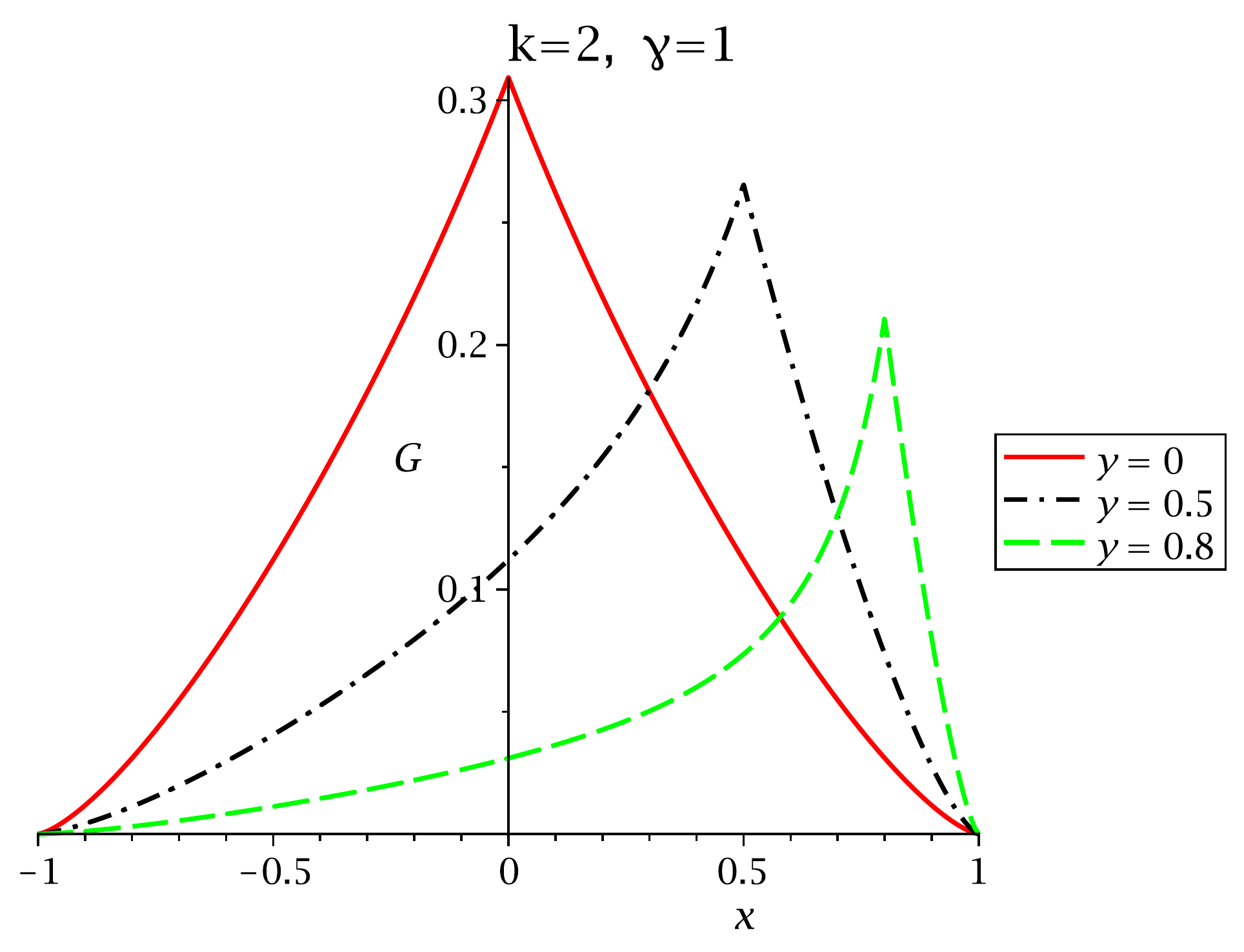}}
\end{minipage}
\caption{Plots of the Green function in the case $m>2$, here $k$ stands for $|\mathbf{k}|$. } 
\label{GreenGen2}  
\end{figure}

\begin{figure}[h] 
\begin{minipage}[h]{0.47\linewidth}
\center{\includegraphics[width=1.0\linewidth]{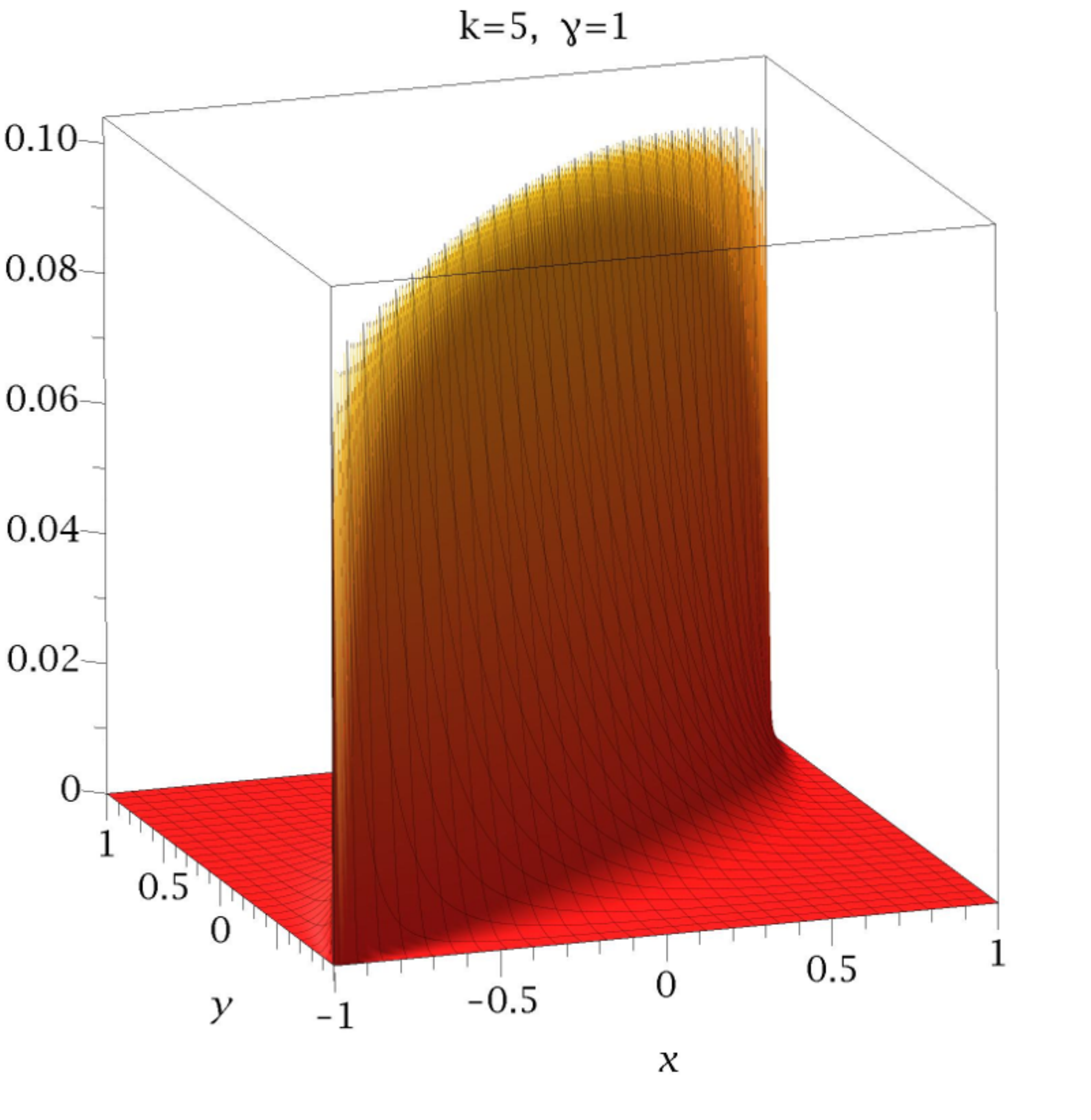}}
\end{minipage}
\hfill
\begin{minipage}[h]{0.47\linewidth}
\center{\includegraphics[width=1.0\linewidth]{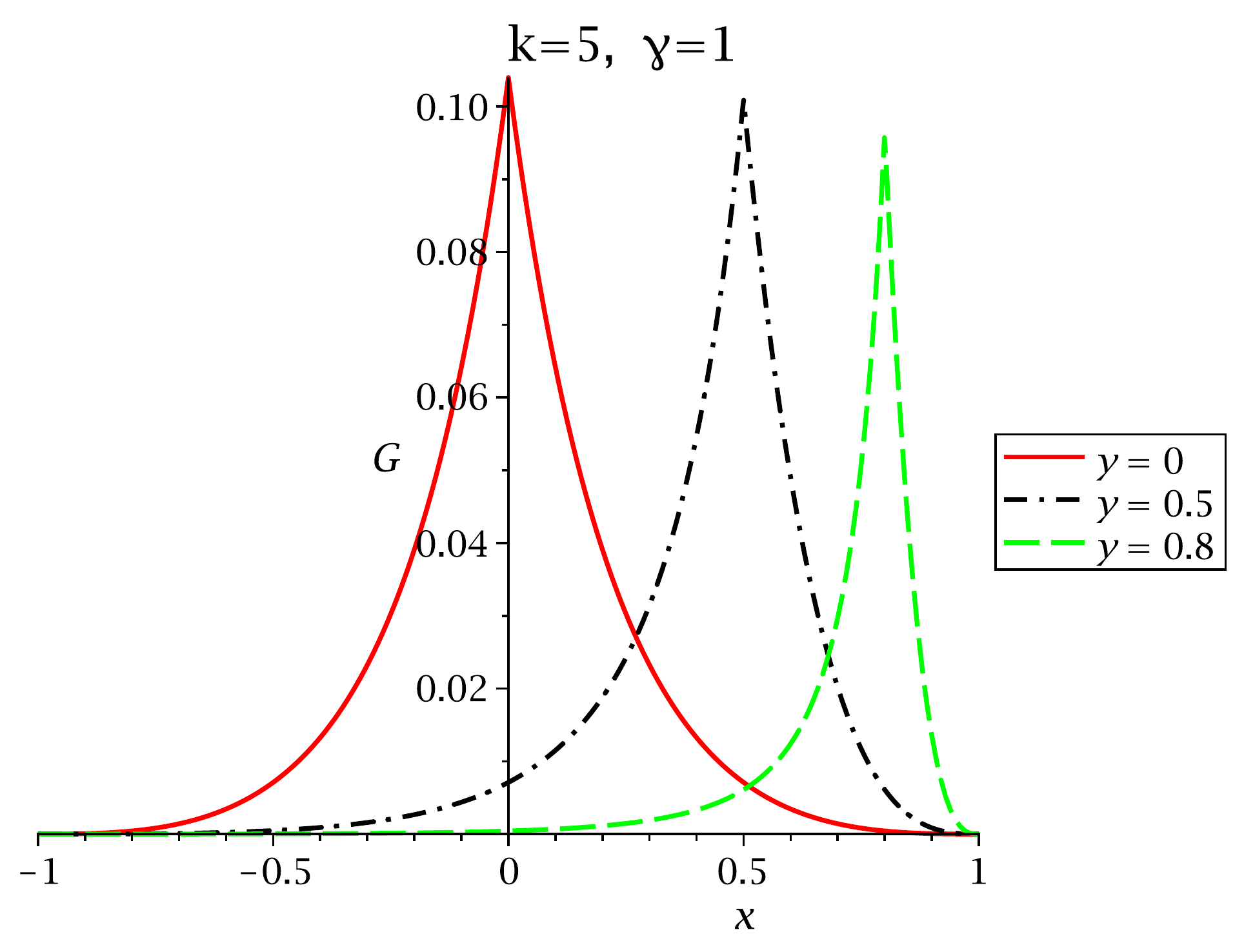}}
\end{minipage}
\caption{Plots of the Green function in the case $m>2$, here $k$ stands for $|\mathbf{k}|$. } 
\label{GreenGen3}  
\end{figure}

\begin{figure}[h]                         
\begin{minipage}[h]{0.47\linewidth}         
\center{\includegraphics[width=1.0\linewidth]{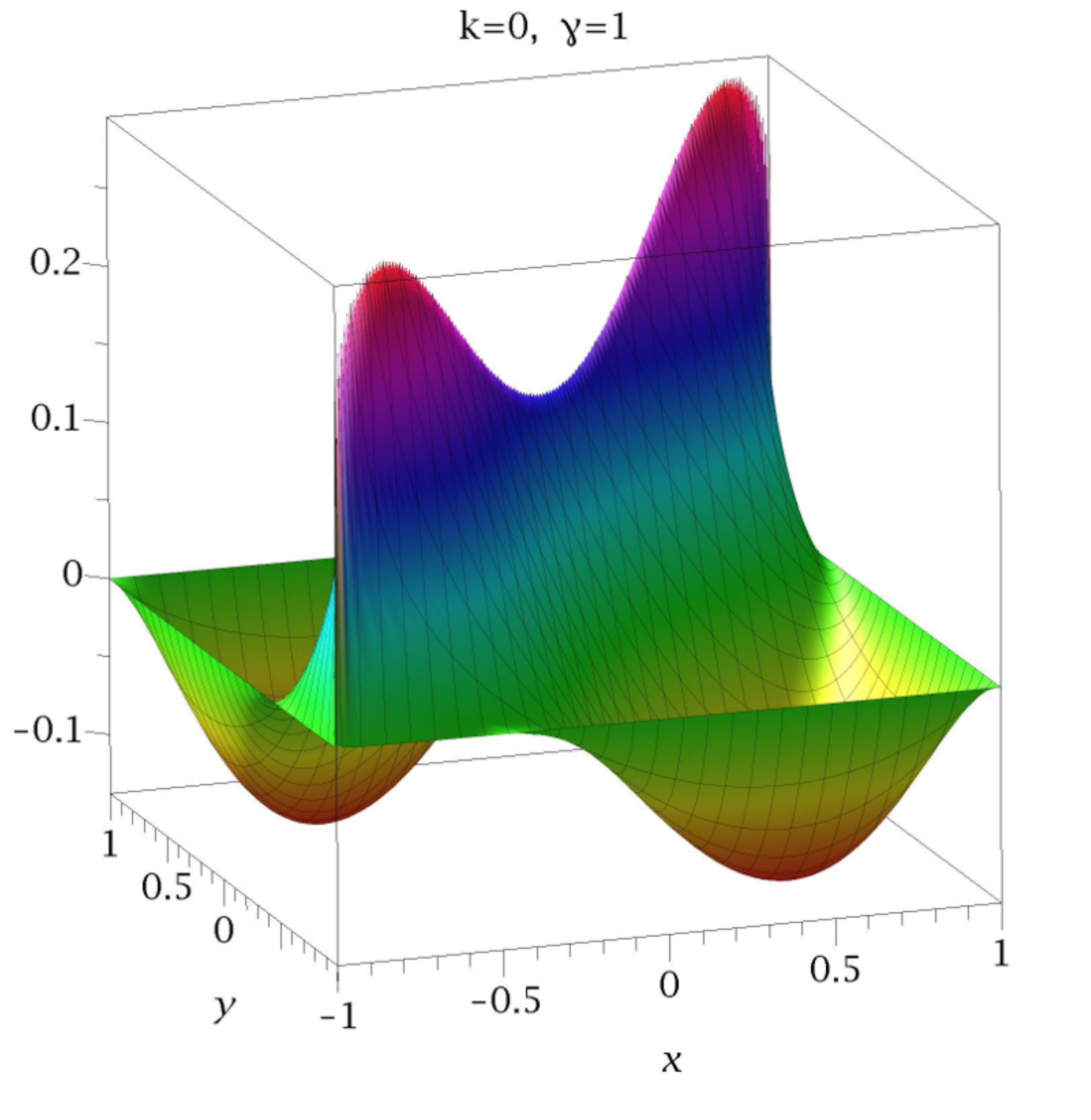}} 
\end{minipage}
\hfill                                      
\begin{minipage}[h]{0.47\linewidth}
\center{\includegraphics[width=1.0\linewidth]{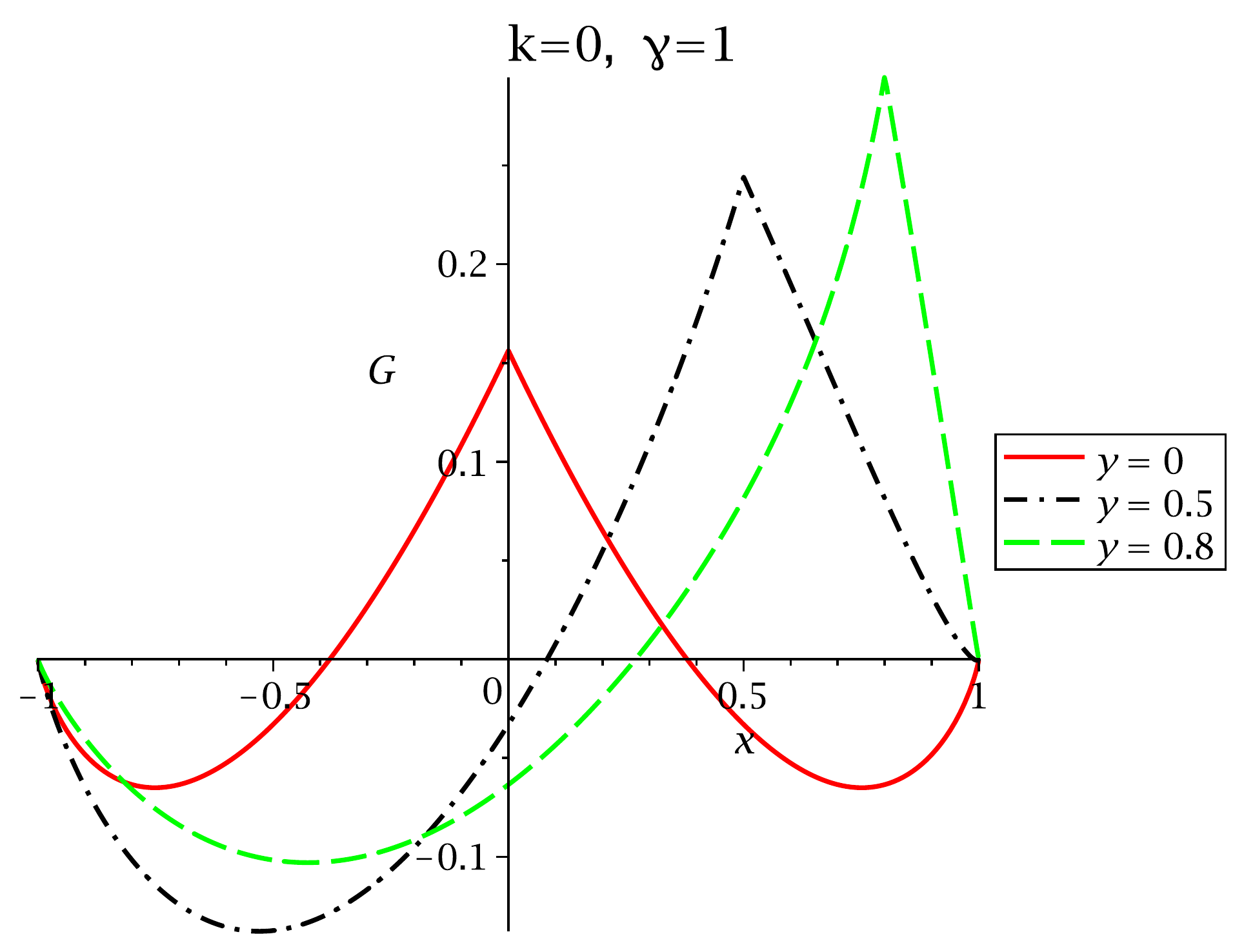}}
\end{minipage}
\caption{Plot of the Green function in the case $m=2$, here $k$ stands for $|\mathbf{k}|$.} \label{GreenParticular}  
\end{figure}
The key ingredient in the Green function method is the expressions for  required Green functions in bounce background. In the case of single scalar field we will need only one such function. Below we give all the necessary details on the solution of the corresponding differential equation and comment on its general properties and special cases.  Here we will closely follow the corresponding section of \cite{VacuumDecaySFT}.

To find a Green function in a bounce background we need to solve the following inhomogeneous Klein-Gordon equation
\begin{equation}
(\triangle^{(4)} + U''(\varphi)) G(\varphi ; x, x') = \delta^{(4)} (x - x')\, .
\end{equation}
To do that, it is convenient to perform Fourier transform with respect to coordinates parallel to the bubble
\begin{equation} 
G(\varphi ; x, x')= \int\frac{d^{3}\mathbf{k}}{(2\pi)^{3}}e^{i\mathbf{k}(\mathbf{z_{\|}}-\mathbf{z_{\|}'})}G(\varphi ; z_{\bot},z_{\bot}',\mathbf{k})
\end{equation}
Then the  Green function $G(\varphi ; z_{\bot},z_{\bot}',\mathbf{k})$ satisfies the equation  ($z, z' = z_{\bot},z'_{\bot}$): 
\begin{equation} 
\label{gen_eq_green_z}
\left[-\frac{d^2}{dz^2}+\mathbf{k}^2+U''(\varphi)\right]G(\varphi ; z,z',\mathbf{k})=\delta(z-z')
\end{equation}
Making change of variables $x=\tanh(\gamma z)$, $y=\tanh(\gamma z')$ we get $\varphi=vx$ ($v^2=12\gamma^2/\lambda$),
\begin{equation} 
U''(\varphi)=-\mu^2+\frac{\lambda}{2}\varphi^2=4\gamma^2-6\gamma^2(1-x^2)
\end{equation}
and the considered equation takes the form
\begin{equation} 
\label{gen_eq_green_x}
\left[\frac{d}{dx}(1-x^2)\frac{d}{dx}-\frac{m^2}{1-x^2}+6\right]G(x,y,\mathbf{k})=-\frac{\delta(x-y)}{\gamma}
\end{equation}
where $m=\frac{1}{\gamma}\sqrt{4\gamma^2+\mathbf{k}^2}$. Here and below to shorten notation we will write $G(x,y,\mathbf{k})$ instead of $G(\varphi ; x,y,\mathbf{k})$. The solution of equation \eqref{gen_eq_green_x} differs depending on whether $m>2$ or $m=2$. In the case  $m>2$ the solution of homogeneous equation is given by 
\begin{equation} 
G(x,y,\bk)=C_1(y)P_2^m(x)+C_2(y)Q_2^m(x)
\end{equation}
where $P_2^m(x)$ and $Q_2^m(x)$ are the associated Legendre functions of the first and second kind, respectively. Accounting for boundary conditions 
\begin{enumerate}
	\item[1)] $G(x,y,\bk)\rightarrow 0$ if $ x,y\rightarrow \pm 1 $
	\item[2)] continuity at $x=y$ 
	\item[3)] jump of the derivative at $x=y$: $\left.\frac{\partial}{\partial{x}}G^{x>y}(x,y,\bk)\right|_{y=x}-\left.\frac{\partial}{\partial{x}}G^{y<x}(x,y,\bk)\right|_{y=x}=-\frac{1}{\gamma(1-x^2)}$
\end{enumerate}
we get
\begin{equation} 
G(x,y,\bk)=\theta(x-y)\frac{\pi}{2\gamma\sin m\pi}P^{-m}_{2}(x)P^{m}_{2}(y)+(x\leftrightarrow y)
\end{equation}
Now, employing the representation of associated Legendre function of the first kind in terms of Jacobi polynomials
\begin{equation}
P_n^m (z) = \left(\frac{z+1}{z-1}\right)^{\frac{m}{2}}(n-m+1)_m P_n^{(-m, m)}(z)
\end{equation} 
and using the fact that for $n=2$ the polynomial expansion of the latter terminates
\begin{equation}
P_2^{(\pm m, \mp m)} (z) = \frac{1}{2} \left[(1\pm m)(2\pm m) - 3 (2\pm m)(1-z) + 3 (1-z)^2\right]
\end{equation}
we obtain \cite{GreenFunctionScalarQFT}:
\begin{multline}
G (x, y, \bk) = \, \frac{1}{2\gamma m} \Bigg\{ \theta (x-y) \left(
\frac{1-x}{1+x}
\right)^{\frac{m}{2}} \left(
\frac{1+y}{1-y}
\right)^{\frac{m}{2}} \left(
1 - 3 \frac{(1-x)(1+m+x)}{(1+m)(2+m)}
\right) \\
 \times \left(
1 - 3 \frac{(1-y)(1-m+y)}{(1-m)(2-m)}
\right) + (x\to y)
\Bigg\}\, . 
\label{GeneralGF1}
\end{multline}
This expression could be also rewritten as:
\begin{multline}
G(x,y,k)=\frac{g^{m/2}(x,y)}{2\gamma m(4-5m^2+m^4)}\left[(1-m^2)^2-3m|x-y|(3xy+1-m^2)+\right.
\\
\left. +9xy(xy-m^2)-3(x^2+y^2)(1-m^2)\right]\, ,
\end{multline}
where
\begin{equation}
\label{Green0_free}  
g(x,y)=\frac{1-|x-y|-xy}{1+|x-y|-xy}
\end{equation}
Here, it is  seen that the obtained Green function is automatically symmetric with respect to $x \leftrightarrow y$ permutations. The plots of this function for different values of m are shown in Figs. \ref{GreenGen1}, \ref{GreenGen2} and  \ref{GreenGen3}. One can see that at large values of m the density of Green function is mostly concentrated near the $x=y$ line.

In the case $m=2$ the differential operator acting on Green function in \eqref{gen_eq_green_x} has zero mode and its inversion is consistently defined only on the subspace of functions orthogonal to this zero mode (Fredholm alternative). The corresponding modified equation for $m=2$ was already considered in \cite{Olejnik,MaximGreenFunctionm2} and is obtained by adding the product of properly normalized zero modes\footnote{The normalization factor $\sqrt{\frac{3\gamma}{4}}$ could be obtained for example from the condition $G(1,1,0)=G_{FV}(1,1,0)$.} ($\varphi_0=\sqrt{\frac{3\gamma}{4}}\frac{1}{\cosh^2(\gamma z)}$) to the right-hand side of the equation \eqref{gen_eq_green_z}, so that
\begin{equation} 
\label{gen_eq_green_z_k0}
\left[-\frac{d^2}{dz^2}+U''(\varphi)\right]G(z,z',0)=\delta(z-z')-\frac{3\gamma}{4\cosh^2(\gamma z)\cosh^2(\gamma z')}
\end{equation}
and equation \eqref{gen_eq_green_x} is changed to
\begin{equation} 
\label{gen_eq_green_x_k0}
\left[\frac{d}{dx}(1-x^2)\frac{d}{dx}-\frac{2^2}{1-x^2}+6\right]G(x,y,0)=\frac{-\delta(x-y)}{\gamma}+\frac{3}{4\gamma}(1-y^2)
\end{equation}
The solution of the latter at $x\ne y$ is then given by
\begin{multline} 
G(x,y,0)=\frac{1}{8\gamma}(1-y^2)\left(1+\frac{1}{1-x^2}\right)+C_1(y)(1-x^2)+
\\
+C_2(y)\left(\frac{3}{4}(1-x^2)\log\frac{1+x}{1-x}+\frac{x}{1-x^2}+\frac{3}{2}x\right)
\end{multline}
or in other form
\begin{equation} 
G(x,y,0)=\frac{1}{8\gamma}(1-y^2)\left(1+\frac{1}{1-x^2}\right)+C_1(y)P_2^2(x)+C_2(y)Q_2^2(x)
\end{equation}
The boundary conditions for $G(x,y,k)$ in this case are given by
\begin{enumerate}
	\item[1)] $G(x,y,k)\rightarrow 0$ if $ x,y\rightarrow \pm 1 $
	\item[2)] continuity at $x=y$
	\item[3)] orthogonality of $G(x,y,0)$ to the zero mode
	\begin{equation} 
	\int\limits_{-\infty}^{\infty}\frac{G(z,z',0)}{\cosh^2(\gamma z)}dz\sim \int\limits_{-1}^1G(x,y,0)dx=0
	\end{equation}
\end{enumerate}
and allow us to fix functions $C_{1,2}(y)$, so that the resulting solution reads 
\begin{multline}
\label{Green0} 
G(x,y,0)=\frac{g(x,y)}{4\gamma}\left\{2-xy+\frac{|x-y|}{4}(11-3xy)+(x-y)^2\right\} \\ +\frac{3}{32\gamma}(1-x^2)(1-y^2)\left(\log g(x,y)-\frac{11}{3}\right)
\end{multline}
with $g(x,y)$ defined in (\ref{Green0_free}).
Note, that this function automatically has the correct discontinuity of the first derivative at $x=y$: $\left.\frac{\partial}{\partial{x}}G^{x>y}(x,y,0)\right|_{y=x}-\left.\frac{\partial}{\partial{x}}G^{y<x}(x,y,0)\right|_{y=x}=-\frac{1}{\gamma(1-x^2)}$ despite the fact that this condition was not explicitly imposed.
The plot of $G(x,y,0)$ could be found in Fig. \ref{GreenParticular}. Similarly in the case of Green function in false vacuum we have
\begin{equation}
(\triangle^{(4)} + U''(v))G_{FV}(x, x') = \delta^{(4)} (x - x')\, .
\end{equation}
The differential operator acting on Green function in false-vacuum does not have zero modes and therefore there is no need to consider the case $m = 2$ separately. Again, performing the Fourier transform with respect to coordinates parallel to the bubble and making change of variables $x=\tanh(\gamma z)$, $y=\tanh(\gamma z')$  we get
\begin{equation} 
\label{gen_eq_green_x_FV}
\left[\frac{d}{dx}(1-x^2)\frac{d}{dx}-\frac{m^2}{1-x^2}\right]G_{FV}(x,y,\mathbf{k})=-\frac{\delta(x-y)}{\gamma}
\end{equation}
The solution of this equation is easy and is given by 
\begin{equation}
\label{green_FV}
G_{FV} (x, y, \bk) = \frac{1}{2\gamma m} \Bigg(\frac{1-|x-y|-xy}{1+|x-y|-xy}
\Bigg)^{\frac{m}{2}} = \frac{g^{m/2}(x,y)}{2\gamma m}
\end{equation}
At infinity the Green function in bounce background behaves just like the Green function in a false vacuum($G_{FV} (z, z', \bk) = G(z, z', \bk)~\text{as}~z, z' \rightarrow \infty$ for any $\bk$) as it should. Transforming the expression (\ref{green_FV}) back to $z$, $z'$ variables we get
\begin{equation} 
\label{ZGreen}
G_{FV}(z,z',\mathbf{k})=\frac{e^{-m(\bk)|z-z'|}}{2 m(\bk)}\, ,\quad m(\bk) = \sqrt{\hat{m}^2+\bk^2}\, , \quad \hat{m}^2 = -\mu^2 + v^2/2 = 4\gamma^2
\end{equation}
Now, if we take the Fourier transforms in variables $z$, $z'$  we recover the usual four-dimensional momentum space propagator:
\begin{equation}
G_{FV}(k) = \frac{1}{k^2 + \hat{m}^2}\, .
\end{equation}
This outcome is of great importance to us as it allows us to use for computations in false vacuum ordinary four-dimensional propagators and still be consistent with the thing wall approximation.

\subsection{Regularization and renormalization}\label{renormalization} 

The next complication in the computation of radiative corrections in quantum field theories is related with the problem of regularization and renormalization of arising ultraviolet divergences.  In order to get finite results for one and two loop radiative corrections to false vacuum decay the general procedure is to introduce necessary counterterms in the lagrangian. For the problem at hand the latter are given by 
\begin{equation}
\LL_{\text{counterterms}} =  \frac{1}{2}\delta\mu^2\Phi^2 + \frac{\delta\lambda}{4!}\Phi^4 + \frac{\delta Z}{2}(\partial_{\mu}\Phi)^2 ,
\label{counterGeneral}
\end{equation} 
where
\begin{equation}
\delta\mu^2=\hbar \delta\mu^2_1+\hbar^2\delta\mu^2_2+... 
\end{equation}
\begin{equation}
\delta\lambda=\hbar \delta\lambda_1+\hbar^2\delta\lambda_2+... 
\end{equation} 
\begin{equation}
\delta Z=\hbar \delta Z_1+\hbar^2\delta Z_2+... 
\end{equation} 
Besides counterterms that will enter the Feynman rules in the computation of particular diagrams at 2-loop order in one of the next subsections, there are also contributions which we will call action counterterms. The latter are introduced by writing leading tree level exponential factor as $\exp{\left(-\frac{1}{\hbar}(S_b+\delta S)\right)}$, where $\delta S=\delta S(\varphi_c)-\delta S(v)$ as the false vacuum decay rate is given by the ratio of partition functions evaluated  around bounce solution and in false vacuum. This action counterterm $\delta S$ could be easily calculated in a general form and is given by
\begin{multline}
\label{countrAction} 
\delta S=\int d^{(4)}x\left(\frac{1}{2}\delta\mu^2(\varphi_c^2-v^2)+\frac{1}{4!}\delta\lambda(\varphi_c^4-v^4)+\frac{1}{2}\delta Z\left(\partial_{\mu}\varphi_c\right)^2\right)=
 \\
=-S_b\left(2\delta Z - \frac{4\delta\lambda}{\lambda}-\frac{3\delta\mu^2}{\gamma^2}\right) 
\end{multline}
The technically most simple strategy at 2-loop level is to use dimensional regularization compared for example with the cut-off regularization. In addition to dimensional regularization we will use two renormalization schemes: Coleman-Weinberg and $\MSbar$. The Coleman-Weinberg  renormalization scheme in our case is defined with the following set of renormalization conditions
\begin{equation}
\label{renormCWconditions}
\left.\frac{\partial^2U}{\partial\Phi^2}\right|_{\Phi=v}=4\gamma^2,\qquad \left.\frac{\partial^4U}{\partial\Phi^4}\right|_{\Phi=v}=\lambda, \qquad \left.\frac{\partial \Gamma_2(p^2)}{\partial p^2}\right|_{p^2=0}=1 ,
\end{equation}
where $\Gamma^{(2)}(p^2)$ is 1PI two-point Green function. The values of mass and coupling counterterms were obtained via renormalization of effective potential, while to obtain wave function renormalization we considered renormalization of two-point correlation function. All these calculations were performed in false vacuum, where as was already mentioned at the end of subsection  \ref{GreenFunctionsubsection} we may use ordinary four-dimensional momentum space propagators. The details of counterterms calculation could be found in appendices  
\ref{Effective potential} and \ref{Wave function renormalization}. At one loop for the action counterterm \eqref{countrAction} we get
\begin{align} 
\label{countrMSbar_1}
\delta S^{(1)}_{\MSbar} &= -\frac{3S_b\lambda}{(4\pi)^2\ep}(4 \pi e^{-\gamma_E})^{\ep} \\
\label{countrCW_1} 
\delta S^{(1)}_{CW} &= -S_b\left(\frac{3\lambda(e^{-\gamma_E} \pi \gamma^{-2})^{\ep}}{16\pi^2}\right)\left(\frac{1}{\ep}+23+\frac{1}{6}\right)
\end{align}
Similarly at 2-loop order we have
\begin{align}
\label{countrMSbar_2} 
\delta S^{(2)}_{\MSbar} &=-\frac{S_b\lambda^2(72-53\ep)}{4(4\pi)^4\ep^2}(4 \pi e^{-\gamma_E})^{2\ep} \\
\label{countrCW_2} 
\delta S^{(2)}_{CW} &=-\frac{144S_b\lambda^2(2e^{-\gE}\pi\gamma^{-2})^{2\ep}}{6144\pi^4}
\left(\frac{1}{\ep^2}+\frac{1}{72 \ep}+\frac{365 \Im\left(\text{Li}_2\left(e^{\frac{i\pi}{3}}\right)\right)}{864 \sqrt{3}}+\frac{233 \pi ^2}{768}-\frac{81313}{1152}\right)
\end{align}

\subsection{Decay rate at one-loop}\label{OneLoopDecay}

To get one-loop expression for false vacuum decay rate we need to evaluate difference of two traces \eqref{funcdet1loop}. To do this we will use the heat kernel method\footnote{See \cite{GreenFunctionScalarQFT} for a similar derivation in a cut-off regularization}\cite{heatkernel-usermanual}. The trace of the inverse of Green function could be then written as 
\begin{equation} 
\tr^{(5)}\ln G^{-1}(\vec{z})=-\int d^{(d-1)}\mathbf{z_{\|}}\int d z_{\bot}\int\limits_0^{\infty}\frac{d\tau}{\tau}K(\vec{z},\vec{z},\tau) ,
\end{equation}
were $\vec{z}=\{z_{\bot},z_{\|}\}$ ($d$-dimensional vector) are the coordinates associated with bubble wall and the heat kernel $K(\vec{z},\vec{z}',\tau)$ is the solution of the heat-flow equation
\begin{equation} 
\frac{\partial}{\partial \tau}K(\vec{z},\vec{z}',\tau) = G^{-1}(\vec{z})K(\vec{z},\vec{z}',\tau)
\end{equation}
satisfying boundary condition $K(\vec{z},\vec{z}',0)=\delta^{(d)}(\vec{z}-\vec{z}')$. Following \cite{GreenFunctionScalarQFT} we perform Laplace transform of the heat kernel
\begin{equation} 
\mathcal{K}(\vec{z},\vec{z}',s)=\int\limits_0^{\infty}d\tau e^{s\tau}K(\vec{z},\vec{z}',\tau)
\end{equation}
and note, that the transformed heat kernel $\mathcal{K}(\vec{z},\vec{z}',s)$ could be treated the same way we treated the Green function in subsection \ref{GreenFunctionsubsection}. Namely, performing Fourier transform with respect to coordinates parallel to the bubble
\begin{equation} 
\mathcal{K}(\vec{z},\vec{z}',s)= \int\frac{d^{3}\mathbf{k}}{(2\pi)^{3}}e^{i\mathbf{k}(\mathbf{z_{\|}}-\mathbf{z_{\|}'})}\mathcal{K}(z_{\bot},z_{\bot}',s,\mathbf{k})
\end{equation}
we see that  $\mathcal{K}(z_{\bot},z_{\bot}',s,\mathbf{k})$ satisfies the equation
\begin{equation} 
\label{gen_eq_kernel_z}
\left[-\frac{d^2}{dz^2}+\mathbf{k}^2+s+U''(\varphi)\right]\mathcal{K}(z_{\bot},z_{\bot}',s,\mathbf{k})=\delta(z-z')
\end{equation}
Comparing equations \eqref{gen_eq_kernel_z} and \eqref{gen_eq_green_z} it is clear that the expression for the transformed kernel $\mathcal{K}(z_{\bot},z_{\bot}',s,\mathbf{k})$ could be obtained from corresponding expression for Green function $G (x, y, \bk)$  \eqref{GeneralGF1} with the substitution $\bk^2 \to \bk^2 + s$. Note, that we don't need to consider the special case $m=2$, as zero modes are already subtracted. Putting all together we get\cite{VacuumDecaySFT}:
\begin{equation} 
I_{\unren}^{(1)}=\frac{1}{2}\int\limits_0^{\infty}\frac{d\tau}{\tau}\int\frac{d^{(d-1)}\mathbf{k}}{(2\pi)^{d-1}}\int d^{(d-1)}\mathbf{z_{\|}}\int d z_{\bot}\mathcal{L}^{-1}_s[G(z_{\bot},z_{\bot},\bk_s)- \\
G_{FV}(z_{\bot},z_{\bot},\bk_s)](\tau) ,
\end{equation}
where $\bk_s$ stands for $\bk^2 + s$, $d=4-2\ep$ and  $\mathcal{L}^{-1}_s$ is the inverse Laplace transform with respect to $s$
\begin{equation}
\mathcal{L}^{-1}_s [F(s)](\tau) = \frac{1}{2\pi i}\int_{\sigma - i\infty}^{\sigma + i\infty} F(s) e^{-s\tau} ds .
\end{equation}
Here $\sigma$ is an arbitrary positive constant, such that the contour of integration lies to the right of all singularities of $F(s)$. All required inverse Laplace transforms could be obtained using shift, scaling and division properties\footnote{See also \cite{GreenFunctionScalarQFT}.}  ($f(\tau) = \mathcal{L}^{-1}_s [F(s)](\tau)$):
\begin{align}
\mathcal{L}^{-1}_s [F(s+b)](\tau) &= e^{-b\tau} f(\tau) , \\
\mathcal{L}^{-1}_s [F(a s)](\tau) &= \frac{1}{a} f (\tau/a) , \\
\mathcal{L}^{-1}_s [s^{-1} F(s)](\tau) &= \int_0^{\tau} d\tau' f(\tau') ,
\end{align}
together with the following elementary transformation 
\begin{equation}
\mathcal{L}^{-1}_s [s^{-z}](\tau) = \frac{\tau^{z-1}}{\Gamma (z)}, \quad \mbox{Re} z > 0
\end{equation}
Taking this way inverse Laplace transforms and performing all integrals except over $\tau$ we get the following expression: 
\begin{equation} 
I_{\unren}^{(1)}=\frac{1}{4\gamma^4}S_b\lambda(4\pi)^{-\frac{d-1}{2}}\int\limits_0^{\infty}d\tau\left[\frac{\gamma}{2}e^{-3\tau\gamma^2}\tau^{5/2+ \ep}\left(\erf(\sqrt{\tau}\gamma)+e^{3\tau\gamma^2}\erf(2\sqrt{\tau}\gamma)\right)\right] ,
\end{equation}
where 
\begin{equation} 
\erf(a)=\frac{2}{\sqrt{\pi}}\int\limits_0^a e^{-t^2}dt
\end{equation}
is the error function. To perform this final integration it is convenient to rewrite error function $\erf(a)$ as a power series 
\begin{equation} 
\erf(a)=\frac{2}{\sqrt{\pi}}e^{-a^2}\sum\limits_{l=0}^{\infty}\frac{2^l a^{2l+1}}{(2l+1)!!}
\end{equation}
and after a term-by-term integration of the above series we get
\begin{equation} 
I_{\unren}^{(1)}=\frac{1}{4\sqrt{\pi}}S_b\lambda(4\pi)^{-\frac{d-1}{2}}\gamma^{-2\ep}\sum\limits_{l=0}^{\infty}\frac{2^{2-2\ep-l}(1+2^{2l+1})\Gamma(l-1+\ep)}{(2l+1)!!}
\end{equation}
The only functions containing $1/\ep$ singularities are $\Gamma(\ep-1)$ and $\Gamma(\ep)$ and thus the pole is generated only by first two terms. The remaining part of the series can be easily summed by putting $\ep=0$
\begin{equation} 
\sum\limits_{l=2}^{\infty}\frac{2^{2-l}(1+2^{2l+1})\Gamma(l-1)}{(2l+1)!!}=\frac{2\pi}{\sqrt{3}}
\end{equation}
Gathering everything together, the unrenormalized one-loop contribution to false vacuum decay rate takes the form
\begin{equation} 
\label{OneLoopSingular}
I_{\unren}^{(1)}=-S_b\left(\frac{3\lambda(e^{-\gamma_E} \pi \gamma^{-2})^{\ep}}{16\pi^2}\right)\left(\frac{1}{\ep}+2-\frac{\pi}{3\sqrt{3}}\right)
\end{equation}
Adding action counterterm corresponding to the Coleman-Weinberg renormalization scheme (\ref{countrCW_1}) the one-loop result in this scheme is then given by
\begin{equation} 
I^{(1)}_{CW}=S_b\left(\frac{3\lambda}{16\pi^2}\right)\left(\frac{\pi}{3\sqrt{3}}+21+\boxed{\frac{1}{6}}\right)
\end{equation}
This result is different from \cite{GreenFunctionScalarQFT} by the framed term $1/6$, which is due to our account for finite wave function renormalization. Numerically
\begin{equation} 
I^{(1)}_{CW}=0.413604\lambda S_b
\end{equation}
Similarly in $\MSbar$ scheme with corresponding action counterterm (\ref{countrMSbar_1}) we get
\begin{equation}
I^{(1)}_{\MSbar} = \frac{3 S_b\lambda}{(4\pi)^2}\left[\frac{\pi}{3\sqrt{3}}-2+\log\left(\frac{4\gamma^2}{\mu_{\MSbar}^2}\right)\right]\, .
\end{equation}

\subsection{Two-loop corrections}\label{TwoLoopDecay}

\begin{figure}[h]
	\center{\includegraphics[width=0.8\textwidth]{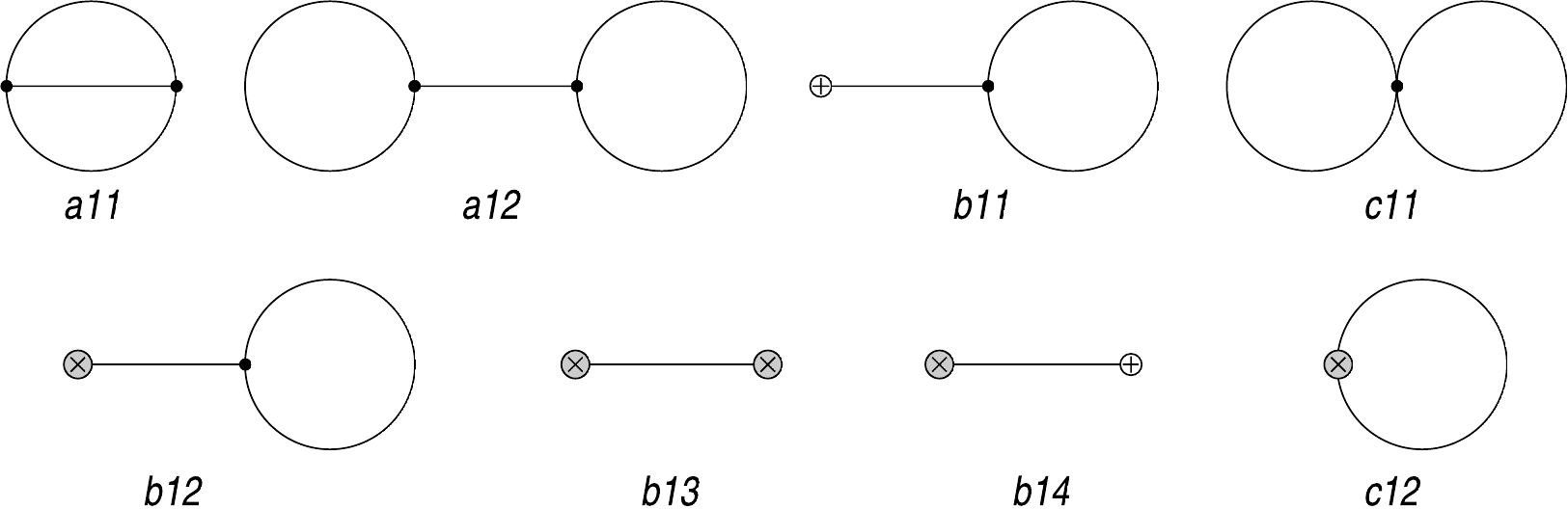}}
	\caption{Two loop Feynman diagrams.}
	\label{FD}
\end{figure}

The evaluation of higher order radiative  corrections to false vacuum decay is similar to quantum mechanical case considered in \cite{VacuumDecayQM}, see also  \cite{Instantons2loop,Olejnik,Instantons3loop,Instantons3loop-SineGordon,QuantumThermalFluctuationsQM} for calculation of other quantities. There are however extra technical complications due to larger spacetime dimension and the need for renormalization. The Feynman rules required for the computation of partition function around bounce solution and in false vacuum are given by
\begin{align}
\vcenter{\hbox{\includegraphics[width=0.08\textwidth]{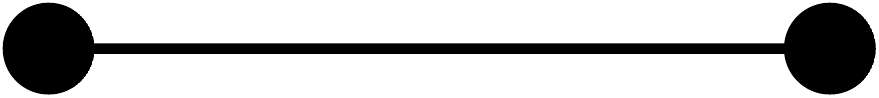}}}=\int\frac{d^{(d-1)}\mathbf{k}}{(2\pi)^{d-1}}e^{i\mathbf{k}(\mathbf{z_{\|}}-\mathbf{z_{\|}'})}G(z_{\bot},z_{\bot}',\mathbf{k})
\label{feynmanroolesGreen}
\end{align}
\begin{align}
\vcenter{\hbox{\includegraphics[width=0.08\textwidth]{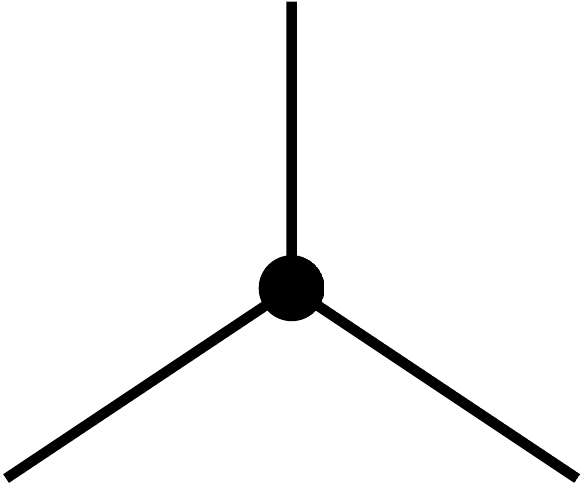}}}=2\gamma\sqrt{3\lambda}x\, ,\qquad &  \vcenter{\hbox{\includegraphics[width=0.08\textwidth]{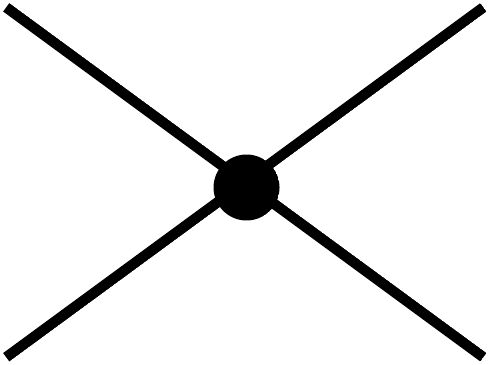}}}=-\lambda, \qquad &\vcenter{\hbox{\includegraphics[width=0.08\textwidth]{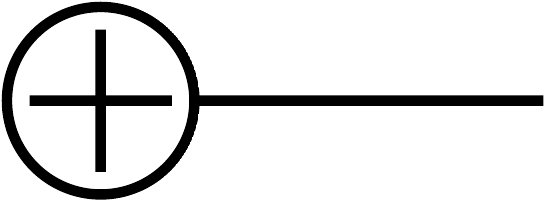}}}=\frac{2x(1-x^2)}{S_b^2} 
\label{feynmanroolesTad}
\end{align}
\begin{align}
\vcenter{\hbox{\includegraphics[width=0.14\textwidth]{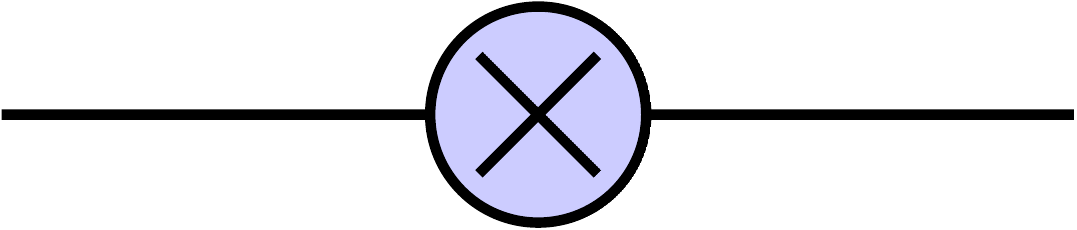}}}=-\delta\mu^2-\frac{\delta\lambda}{2}v^2x^2+2\gamma^2(3x^2-1)\delta Z\,
\label{feynmanroolesTad2}  
\end{align}
\begin{align}
\vcenter{\hbox{\includegraphics[width=0.08\textwidth]{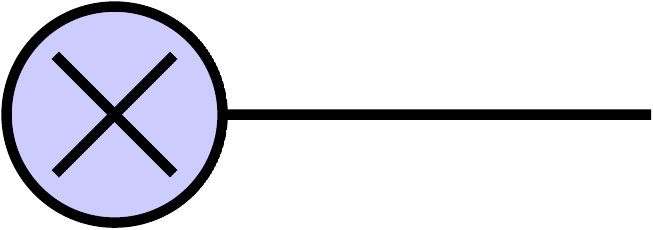}}}=vx\left(\delta\mu^2+\frac{\delta\lambda}{3!}v^2x^2-2\gamma^2\delta Z(2x^2-1)\right), 
\label{feynmanroolesTad3}
\end{align}
where $x=\tanh(\gamma z)$. The vertex with a plus sign inside a circle is a tadpole vertex coming from integration measure\footnote{See \cite{VacuumDecayQM} for its derivation in the case of quantum mechanics.}, while all other vertexes come from lagrangian. To derive the expression for tadpole vertex it is convenient to use the same Faddeev-Popov trick as in the case of quantum mechanics\cite{Instantons2loop,Olejnik,VacuumDecayQM}. First, we should note that in this case we have 4 zero modes related to coordinates of the center of the bounce $x_{0}$. The latter are given by
\begin{equation}
\varphi_{i} = S_b^{-1/2}\partial_{i}\varphi \, ,\quad i = 1\ldots 4.
\end{equation} 
where $ \varphi$ is the bounce solution as before. Now, writing unity decomposition as
\begin{equation}
1 = \int d^4 x_{0} \prod_{i=1}^4 \delta (f_i (x)) \frac{\partial f_i (x)}{\partial x_i}
\end{equation}
where ($\Phi$ is the original field from the lagrangian)
\begin{equation}
f_i = \la \Phi - \varphi | \varphi_i\ra = \la \phi_{qu} |  \varphi_i\ra, 
\end{equation}
and evaluating $\frac{\partial f_i (x)}{\partial x_i}$ we get
\begin{align}
1 =& \int d^4 x_{0} \prod_{i=1}^4  \left(-S_b^{1/2} + \la \phi_{qu} | \frac{\partial\varphi_i}{\partial x_i}\ra\right)\delta (\xi_{0,i}) \nonumber \\
=& \int d^4 x_{0} \prod_{i=1}^4  \left(-S_b^{1/2} + \int d^4 x \phi \frac{\partial\varphi_i}{\partial x_i}\ra\right)\delta (\xi_{0,i})
\end{align}
where $\xi_{0,i}$ are expansions coefficients of $\phi_{qu}$ in front of zero mode eigenstates. Performing field rescaling $\phi\to \hbar^{1/2}\phi$ the required vertex at two loops takes the form
\begin{equation}
-\hbar S_b^{-2} \int d^4 x \phi\triangle \varphi
\end{equation}
At two-loop lebel we need to calculate diagrams presented in Fig. \ref{FD}

All diagrams except sunset diagram $a11$ could be calculated more or less straightforwardly. Take for example diagram $c11$.  Using Feynman rules \eqref{feynmanroolesGreen}-\eqref{feynmanroolesTad3} we have
\begin{multline} 
I_{c11}=-\frac{\lambda}{8}\int d^{(d-1)}\mathbf{z_{\|}}\int d z_{\bot}\int\frac{d^{d-1}\mathbf{k_1}}{(2\pi)^{d-1}}\int\frac{d^{d-1}\mathbf{k_2}}{(2\pi)^{d-1}}(G(z_{\bot},z_{\bot},\mathbf{k_1})G(z_{\bot},z_{\bot},\mathbf{k_2})- \nonumber \\
-G_{FV}(z_{\bot},z_{\bot},\mathbf{k_1})G_{FV}(z_{\bot},z_{\bot},\mathbf{k_2}))
\end{multline}
Note, that the integrand of this expression consist from two parts corresponding to  bounce solution and false vacuum. It is due to the fact, that false vacuum decay rate in path integral formulation is given by the ratio of partition sums evaluated around bounce and false vacuum solutions. The same is true for all diagrams except ones with tadpole vertex because the latter comes from the integration measure at bounce solution and has no counterpart for false vacuum. Going from $z_{\bot}$ to $x=\tanh (\gamma z_{\bot})$ variable and making use of the formula (\ref{volume}) the expression for $I_{c11}$ takes the form
\begin{equation}
I_{c11}=-S_b\left(\frac{\lambda^2}{32\gamma^4}\right)\int\limits_{-1}^{1}\frac{dx}{1-x^2}\left(G^2(x)-G_{FV}^2 (x)\right)
\end{equation}
Finally, using the expressions for integrals from Green functions
\begin{multline}
G (x) = \int\frac{d^{d-1}\bk}{(2\pi)^{d-1}} G(z_{\bot}, z_{\bot}, \bk) = \frac{\gamma^{2-2\ep} e^{-\ep\gE}}{8\pi^{2-\ep}}\Bigg\{ \frac{1-3 x^2}{\ep} + 4 + x^2 (-6 + \sqrt{3}\pi (x^2-1)) \\
+ \ep \Big[ 10 - 12 x^2 + \frac{\pi^2}{12}(1-3 x^2) - \frac{\sqrt{3}\pi}{2} x^2 (x^2-1) (\log 3 - 4) \\
+ 3 i\sqrt{3} x^2 (x^2 - 1)\Big\{ \Li2 \Big(\frac{3-i\sqrt{3}}{6}\Big) - \Li2 \Big(\frac{3+i\sqrt{3}}{6}\Big)\Big\}\Big]\Bigg\}
\end{multline}
and
\begin{equation}
G_{FV} (x) = \int\frac{d^{d-1}\bk}{(2\pi)^{d-1}} G_{FV}(z_{\bot}, z_{\bot}, \bk) = G (1) = -\frac{\gamma^{2-2\ep}e^{-\ep\gE}}{4\pi^{2-\ep}}\left\{
\frac{1}{\ep} + 1 + \ep \left(1+\frac{\pi^2}{12}\right)
\right\}
\end{equation}
we get
\begin{multline} 
I_{c11}
=\frac{S_b\lambda^2(e^{-\gamma_E} \pi \gamma^{-2})^{2\ep}}{32\pi^4}\left[\frac{3}{16}\left(\frac{1}{\ep^2}+\frac{2}{\ep}+2\right)-\frac{\pi}{20\sqrt{3}\ep}-\frac{\sqrt{3}\pi}{40}+\frac{29\pi^2}{1120}+\right.
\\ 
\left.+\frac{1}{40\sqrt{3}}\left\{\pi\log 3-6i \Li2 \left(\frac{3-i\sqrt{3}}{6}\right)+6i \Li2 \left(\frac{3+i\sqrt{3}}{6}\right)\right\}\right]
\end{multline}
As a second example lets take the $b11$ diagram with a tadpole vertex. 
\begin{multline}
I_{b11}=\frac{1}{2}\int d z_{\bot}\int d z'_{\bot}\int d^{(d-1)}\mathbf{z_{\|}}\int d^{(d-1)}\mathbf{z'_{\|}}\int\frac{d^{d-1}\mathbf{k_1}}{(2\pi)^{d-1}}\int\frac{d^{d-1}\mathbf{k_2}}{(2\pi)^{d-1}}e^{i\mathbf{k_1 (z_{\|}-z'_{\|})}}\times
 \\
\times\left\{V_t(z_{\bot})V_3(z'_{\bot})G(z_{\bot},z'_{\bot},\mathbf{k_1})G(z'_{\bot},z'_{\bot},\mathbf{k_2})\right\}
\end{multline}

Going from $z_{\bot}$, $z'_{\bot}$ to $x=\tanh (\gamma z_{\bot})$, $y=\tanh (\gamma z'_{\bot})$ variables  and performing integrations over coordinates parallel to the bubble the expression for $I_{c11}$ takes the form
\begin{equation}
I_{b11}= \frac{2\sqrt{3\lambda}\lambda}{4\gamma^4S_b}\int\limits_{-1}^1\int\limits_{-1}^1\frac{xydxdy}{(1-y^2)}G(x,y,0)G(y) ,
\end{equation}
which is easily integrated and we get
\begin{equation}
I_{b11}=-\frac{2\sqrt{3\lambda}\lambda(e^{-\gamma_E} \pi \gamma^{-2})^{\ep}}{128\pi^2S_b}\left(\frac{1}{\ep}+\frac{7\sqrt{3}\pi}{45}\right)
\end{equation}
This example shows the use of the expression for Green function  $G(x,y,0)$ in the  special case with $m=2$, see section \ref{GreenFunctionsubsection}\footnote{In \cite{GreenFunctionScalarQFT} this problem was solved in another way by dividing the Green function for general $m$ \eqref{GeneralGF} into  odd and even parts. Then, it turned out that the odd part containing infinity cancels out in actual calculations.  Both these methods give the same result. Nevertheless we believe that our method is more mathematically consistent as in this case Green function does not contain any unnatural singularities from the very beginning.}. We see that the diagram $I_{b11}$ with a tadpole vertex differs from the diagram $I_{c11}$ containing only    
vertexes from lagrangian by a factor $\sqrt{\lambda} S_b^2$. In the thing wall approximation the bounce action is large $S_b \sim R^3$ and as a consequence the diagram with tadpole vertex $I_{b11}$ is heavily suppressed. The same is true for diagram $I_{b14}$, while all other diagrams have the same scaling $\sim R^3$. So, the diagrams with tadpole vertex from integration measure do not contribute to the decay rate in a thin wall approximation considered in this paper. It should be noted, that this is completely different from  the quantum mechanical case, in which  the largest contribution to the vacuum decay rate was instead given by diagrams containing similar tadpole vertex\cite{Instantons3loop,VacuumDecayQM}. The values of all other diagrams could be found in a mathematica files accompanying this article and the details of sunset diagram evaluation could be found in appendix \ref{sunset-appendix}.

Summing contribution of all diagrams and adding two-loop action counterterms (\eqref{countrCW_2} or \eqref{countrMSbar_2})
we finally get\footnote{Note, that the equation (\ref{decay_final1}) should be considered as the preferred definition for the false vacuum decay rate, see  \cite{Instantons2loop,Instantons3loop,Instantons3loop-SineGordon,VacuumDecayQM} for more details. On the other hand, the form (\ref{decay_final2}) is more convenient for carrying out the renormalization procedure.} 
\begin{equation}
\label{decay_final1}  
\frac{\Gamma}{V}=\left(\frac{S_b}{2\pi\hbar}\right)^2\frac{(2\gamma)^5R}{\sqrt{3}}\exp(-\frac{1}{\hbar}S_b+ I^{(1)})\left\{1+\hbar I^{(2)}+\mathcal{O}(\hbar^2)\right\}
\end{equation}
or
\begin{equation}
\label{decay_final2}  
\frac{\Gamma}{V}=\left(\frac{S_b}{2\pi\hbar}\right)^2\frac{(2\gamma)^5R}{\sqrt{3}}\exp\big(-\frac{1}{\hbar}S_b + I^{(1)} + \hbar I^{(2)} +\mathcal{O}(\hbar^2)\big)\, , 
\end{equation}
where $I^{(2)}$ is finite full two loop correction, which in $\MSbar$ and $CW$ schemes are given by 
\begin{multline}
I^{(2)}_{\MSbar} = \frac{S_b\lambda^2}{8\pi^4}\left[-\frac{3}{4}+\frac{7\sqrt{3}\pi}{160} - \frac{197\pi^2}{8960}-\frac{142-3\sqrt{3}\pi}{384}\log\left(\frac{4\gamma^2}{\mu_{\MSbar}^2}\right)+\frac{9}{256}\log^2\left(\frac{4\gamma^2}{\mu_{\MSbar}^2}\right)\right.
 \\
-\left.\frac{3\sqrt{3}}{320}\left\{\pi\log 3-6i \Li2 \left(\frac{3-i\sqrt{3}}{6}\right)+6i \Li2 \left(\frac{3+i\sqrt{3}}{6}\right)\right\}+s_0\right]
\end{multline}
\begin{multline}
I^{(2)}_{CW}=\frac{\lambda^2S_b}{8\pi^4}\Bigg[-\frac{9}{512} \sqrt{3} \Im\left(\text{Li}_2\left(e^{\frac{i \pi }{3}}\right)\right)+\frac{19 \Im\left(\text{Li}_2\left(e^{\frac{i \pi }{3}}\right)\right)}{144 \sqrt{3}}-\frac{9}{160} i \sqrt{3} \text{Li}_2\left(\frac{3+i\sqrt{3}}{6}\right)+\\
   +\frac{9}{160} i \sqrt{3} \text{Li}_2\left(\frac{3-i\sqrt{3}}{6}\right)+s_0+\frac{5003 \pi ^2}{143360}-\frac{47 \sqrt{3} \pi
   }{256}-\frac{73843}{6144}-\frac{3}{320} \sqrt{3} \pi  \log (3) \Bigg]
\end{multline}
Here $s_0$ is the finite part of the sunset diagram $a11$, see appendix  \ref{sunset-appendix}. Numerically 
\begin{equation} 
I^{(2)}_{CW}=-0.015267 \lambda S_b
\end{equation}
Here, we would like to note that part of two-loop corrections related to bounce renormalization  considered in \cite{GreenFunctionScalarQFT} is contained in the sum of our diagrams $a12$, $b12$ and $b13$. These diagrams are obtained by inserting tadpole contribution for scalar field self-energy found in \cite{GreenFunctionScalarQFT} into corresponding one-loop diagram for false vacuum decay rate.
\begin{equation}
I_{a12}+I_{b12}+I_{b13}=\frac{3S_b\lambda^2}{256\pi^4}\left(\frac{291}{8}-\frac{37\pi}{4\sqrt{3}}+\frac{5\pi^2}{168}-\boxed{\left(\frac{189}{32}+\frac{11\pi}{120\sqrt{3}}\right)}\right)
\end{equation}
Numerically, this contribution is of the order of 10\% of full two-loop result.
\begin{figure}[h]
	\center{\includegraphics[width=0.5\textwidth]{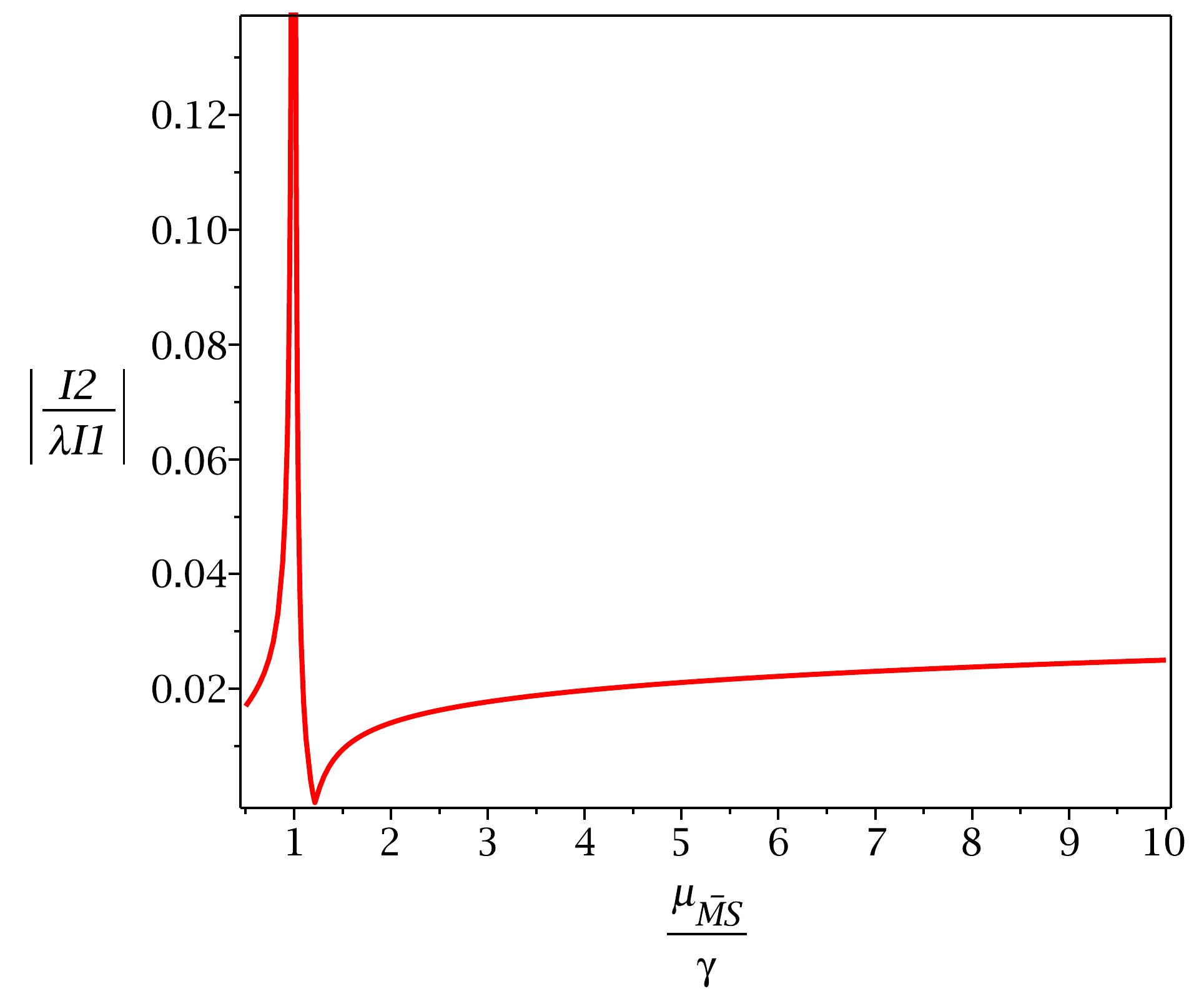}}
	\caption{The ratio of $I_{\MSbar}^{(2)}$ to $\lambda I_{\MSbar}^{(1)}$ as a function of $\mu_{\MSbar}/\gamma$.}
	\label{ratio}
\end{figure} 

Finally, lets discuss the scale dependence of vacuum decay rate in $\MSbar$ scheme. The decay rate is a measurable quantity and as such it is independent from the scale choice. The latter means that our result in $\MSbar$ scheme need to satisfy for example the following constraint:
\begin{equation}
\frac{d}{d \log \mu_{\MSbar}}\left(-\frac{1}{\hbar} S_b + I_{\MSbar}^{(1)} + \mathcal{O}(\hbar)\right)=0
\end{equation}
It is easy to verify, that it is indeed the case by expressing $S_b$ and $I_{\MSbar}^{(1)}$ in terms of the parameters $\gamma,~g,~\lambda$ of the lagrangian and applying to them the renormalization group equations \cite{SMtunneling4}:
\begin{equation}
\frac{d\gamma}{d \log \mu_{\MSbar}}=\hbar\ \frac{\gamma\lambda}{32\pi^2}
\end{equation}
\begin{equation}
\frac{dg}{d \log\mu_{\MSbar}}=\hbar\ \frac{3g\lambda}{16\pi^2}
\end{equation}
\begin{equation}
\frac{d\lambda}{d \log\mu_{\MSbar}}=\hbar\ \frac{3\lambda^2}{16\pi^2}
\end{equation}
Note, that in the thing wall approximation ($R\to\infty$, $g\to 0$) we need to leave in renormalization group equations only leading order terms in coupling $g$. Moreover, it turns out  that within thin-wall approximation the knowledge of one-loop beta-functions is sufficient to check the coefficients in front of logarithms with $\mu_{\MSbar}$ dependence at two-loop order also \cite{VacuumDecaySFT}. Thus, we may conclude that our method is free from renormalization-scale uncertainty\footnote{As usual there may be some residual scale dependence left at one perturbation order higher.}. To understand the qualitative significance of the result in $\MSbar$ scheme lets plot the ratio of $I_{\MSbar}^{(2)}$ to $\lambda I_{\MSbar}^{(1)}$ as a function of $\mu_{\MSbar}/\gamma$. Fig. \ref{ratio} shows this ratio for a reasonable values of $\mu_{\MSbar} \gtrsim 2\gamma$. In the region, where neither $I_{\MSbar}^{(1)}$ nor $I_{\MSbar}^{(2)}$ are close to zero, this ratio is about $0.02-0.03$. So, we see that two loop corrections are small and could be safely neglected whenever we do not interested in accuracy greater than $2-4\%$.

\section{Conclusion}\label{Conclusion}

In this work we presented detail calculation of two-loop radiative corrections to false vacuum decay in a scalar field theory with cubic and quartic self-interactions. To be able to get analytical results we used thin wall approximation and dimensional regularization. The results presented in two different schemes: Coleman-Weinberg and $\MSbar$. It is shown that the two-loop contribution is of order $2-4\%$ of one-loop result in both schemes and as such could be safely neglected if we are not interested in this type of accuracy.  

There several possible future directions and questions which were not touched in this paper. First, it would be interesting to study in more detail the different contributions, such as bounce renormalization and so on, to see their relative importance in complete two-loop result. Second, it would be certainly interesting to consider vacuum decay in theories containing fermions and gauge bosons at 2-loop level, first of course within thin wall approximation. Third, it certainly would be nice to go beyond thin wall approximation. The first steps in this direction were done at one-loop level in \cite{beyondThinWall1,beyondThinWall2}. Unfortunately, the mentioned studies used Gel'fand-Yaglom method and we would like to have the same result in Green function method to have possibility to generalize it beyond one-loop. In this respect it also seems worth studying the possibility to study the false vacuum decay with the methods of lattice field theory and lattice perturbation theory in bounce background and at false vacuum.     

\section*{Acknowledgements}

The authors would like to thank O.~Veretin, A.~Bednyakov and V.~Braguta for interesting and stimulating discussions. This work was supported by Foundation for the Advancement of Theoretical Physics and Mathematics "BASIS". 

\appendix

\section{Effective potential}
\label{Effective potential}
\begin{figure}[hh]
	\center{\includegraphics[width=0.67\textwidth]{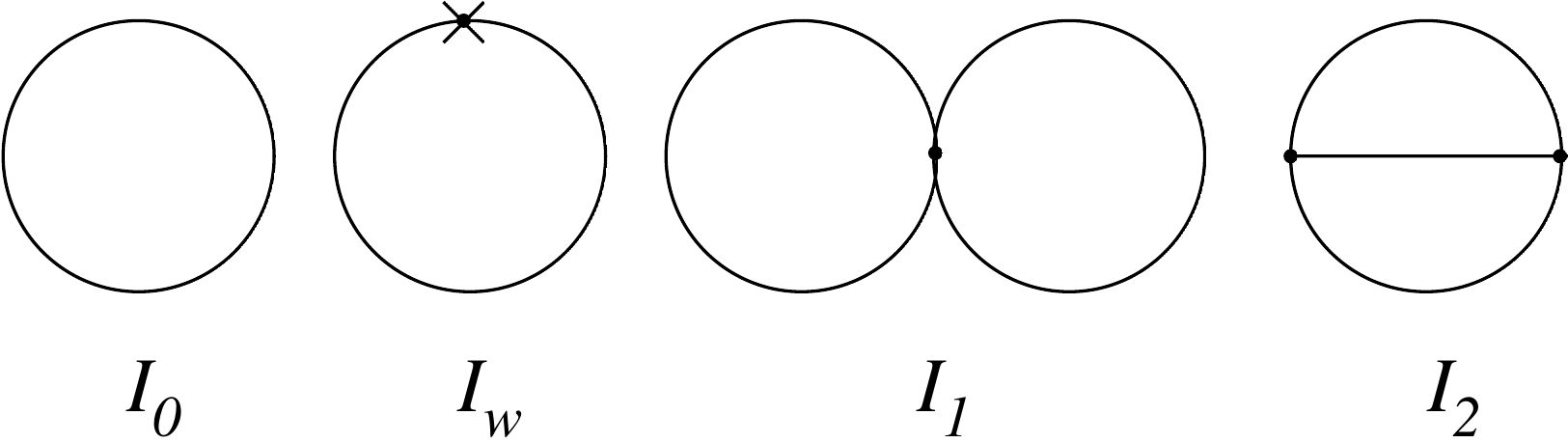}}
	\caption{Diagrams contributing to effective potential. $I_0$ is one loop contribution and $I_w, I_1, I_2$ are two loop contributions.}
	\label{EffectivePotential}
\end{figure}

To calculate effective potential the most convenient way is to consider one-particle irreducible (1PI) effective action \cite{Jackiw}. The latter is given by Legendre transform of partition function
\begin{equation}
\Gamma [\phi] = -\hbar \ln Z[J] + \int d^4 x J(x)\phi (x) ,
\end{equation}
so that the source $J(x)$ is the solution of
\begin{equation}
\phi (x) = \hbar \frac{\delta\ln Z[J]}{\delta J(x)} .
\end{equation}
Now expanding around classical background solution $\hat{\varphi}$ we get ($\Phi = \hat{\varphi} + \hbar^{1/2}\hat\Phi$):
\begin{equation}
e^{-\frac{1}{\hbar} \left(\Gamma[\phi]-\int d^4 x J(x)\phi(x)\right)} = Z[J] 
= \int [d\hat\Phi] e^{-\frac{1}{\hbar}\left(S[\hat{\varphi}]-\int d^4 x J(x)\hat{\varphi}(x)\right) - \frac{1}{2}\int d^4 x \hat\Phi G^{-1}(\hat{\varphi})\hat\Phi} e^{-\frac{1}{\hbar}\int d^4 x\mathcal{L}_I (\hat{\varphi}, \hat\Phi)} ,
\end{equation}
where $\mathcal{L}_I (\hat{\varphi}, \hat\Phi)$ is the interaction lagrangian. Thus, up to 2-loop order the effective potential contains three pieces:  tree level potential plus one-loop correction given by functional determinant of $ G^{-1}(\hat{\phi})$ and minus two-loop diagrams.  Diagrams contributing at one and two loop levels to effective potential are shown in Fig. \ref{EffectivePotential}.
At one loop order to calculate functional determinant it is convenient to use heat kernel method, which we used to obtain one-loop expression for decay rate. This way for unrenormalized one-loop effective potential we get ($d=4-2\ep$)
\begin{multline} 
V^{(1)}(\hat{\varphi})= \frac{\hbar}{2}\frac{1}{VT}\ln\det G_{FV}^{-1}(x) = \frac{\hbar}{2}\frac{1}{VT}\tr\ln G_{FV}^{-1}(x)
\\
= -\frac{\hbar\pi^{\frac{d}{2}-1}}{2\Gamma\left(\frac{d-1}{2}\right)(2\pi)^{d-1}}\int\limits_0^{\infty}k^{d-2}dk\int\limits_0^{\infty}\frac{d\tau}{\tau^{3/2}}(e^{-(k^2+\hat{m}^2)\tau}-e^{-k^2\tau}) \\
= -\frac{\hbar}{32\pi^2} (4\pi)^{\ep} e^{\ep\gamma_E} (\hat{m}^2)^{2-\ep} \left[
\frac{1}{2\ep} + \frac{3}{4}
\right] .
\end{multline}
where $\hat{m}^2=\frac{\lambda}{2}\hat{\varphi}^2-\mu^2$. To perform renormalization we introduce counterterms as in \eqref{counterGeneral}. In this paper we are using two different renormalization schemes: Coleman-Weinberg \eqref{renormCWconditions} and $\MSbar$. From renormalization of effective potential in Coleman-Weinberg scheme we have
\begin{align} 
(\delta\mu^2)^{(1)}_{CW} &= \frac{\lambda\gamma^2(e^{-\gamma_E} \pi \gamma^{-2})^{\ep}}{16\pi^2}\left(-\frac{1}{\ep}+29\right) \\
\delta\lambda^{(1)}_{CW} &= -\frac{3\lambda^2(e^{-\gamma_E} \pi \gamma^{-2})^{\ep}}{32\pi^2}\left(-\frac{1}{\ep}+3\right)
\end{align}
Similarly in $\MSbar$ scheme
\begin{equation} 
(\delta\mu^2)^{(1)}_{\MSbar} =-\frac{\lambda\mu^2}{2(4\pi)^2\ep}(4 \pi e^{-\gamma_E})^{\ep},\qquad
\delta\lambda^{(1)}_{\MSbar}=\frac{3\lambda^2}{2(4\pi)^2\ep}(4 \pi e^{-\gamma_E})^{\ep}
\end{equation}
and the renormalized effective potential in $\MSbar$ scheme takes the form 
\begin{equation}
V^{(1)}(\hat{\varphi})=\frac{\hbar\hat{m}^4}{64\pi^2}\left(\log\left(\frac{\hat{m}^2}{\mu_{\MSbar}^2}\right)-\frac{3}{2}\right)
\end{equation}
At 2-loop level we need to calculate three Feynman diagrams shown in Fig. \ref{EffectivePotential}. Their calculation is easy and reduces to the calculation of single-scale one and two-loop tadpole diagrams. As a result we get
\begin{align}
I_1 &= -\frac{\hbar^2\lambda\hat{m}^4(2e^{-\gE}\pi\hat{m}^{-2})^{2\ep}}{2048\pi^4}\left(\frac{1}{\ep^2}+\frac{2}{\ep}+3+\frac{\pi^2}{6}\right) \\
I_2 &= -\frac{\hbar^2\lambda^2\hat{m}^2\hat{\varphi}^2(2e^{-\gE}\pi\hat{m}^{-2})^{2\ep}}{2048\pi^4}\left(\frac{1}{\ep^2}+\frac{3}{\ep}+I_{2, const}\right) \\
I_w &= -\frac{\hbar^2\hat{m}^2\left(\hat{m}^2\delta Z^{(1)}-(\delta\mu^2)^{(1)}-\frac{\delta\lambda^{(1)}}{2}\hat{\varphi}^2\right)(2e^{-\gE}\pi\hat{m}^{-2})^{\ep}}{32\pi^2}\left(\frac{1}{\ep}+1\right) ,
\end{align}
where
\begin{equation}
I_{2, const}=\frac{1}{64} \left(8 \sqrt{3} \Im\left(\text{Li}_2\left(e^{\frac{i \pi }{3}}\right)\right)-42-\pi ^2\right)
\end{equation}
Renormalizing 2-loop effective potential we then obtain
\begin{align}
(\delta\mu^2)^{(2)}_{CW} &= \frac{(e^{-\gE}\pi\gamma^{-2})^{2\ep}\lambda^2\gamma^2}{256\pi^4}\left(-\frac{1}{\ep^2}+\frac{19}{\ep}+\frac{1}{4}\left(-847+2I_{2, const}-\pi^2\right)\right) \\
\delta\lambda^{(2)}_{CW} &= \frac{(e^{-\gE}\pi\gamma^{-2})^{2\ep}\lambda^3}{1024\pi^4}\left(\frac{9}{\ep^2}-\frac{57}{\ep}-\frac{21}{2}-6I_{2, const}+\frac{5}{2}\left(90+\pi^2\right)\right)
\end{align}
and
\begin{equation} 
(\delta\mu^2)^{(2)}_{\MSbar}=-\frac{\lambda^2\mu^2(2-\ep)}{4(4\pi)^4\ep^2}(4 \pi e^{-\gamma_E})^{2\ep},\qquad
\delta\lambda^{(2)}_{\MSbar}=\frac{3\lambda^3(3-2\ep)}{4(4\pi)^4\ep^2}(4 \pi e^{-\gamma_E})^{2\ep} ,
\end{equation}
which are in agreement with previously obtained results \cite{FunctionalMethodsPerturbationTheory,ScalingBehaviorPhi4,AnApproachEffectivePotential}. For renormalized 2-loop effective potential in $\MSbar$ scheme we got
\begin{multline}
V_2(\hat{\varphi})=\frac{\hbar^2\lambda}{3(8\pi)^4}\Bigg(\frac{3}{2}\left(3\lambda^2\hat{\varphi}^4-8\lambda\mu^2\hat{\varphi}^2+4\mu^4\right)\log^2\left(\frac{\hat{m}^2}{\mu_{\MSbar}^2}\right)-\\
-3\left(5\lambda^2\hat{\varphi}^4-12\lambda\mu^2\hat{\varphi}^2+4\mu^4\right)\log\left(\frac{\hat{m}^2}{\mu_{\MSbar}^2}\right)-\\-\hat{m}^2\left(3\lambda\hat{\varphi}^2+6\mu^2+\lambda\hat{\varphi}^2\left(\pi^2+6-6I_{2, const}\right)\right)\Bigg)
\end{multline}

\section{Wave function renormalization}
\label{Wave function renormalization}

\begin{figure}[h]
	\center{\includegraphics[width=0.37\textwidth]{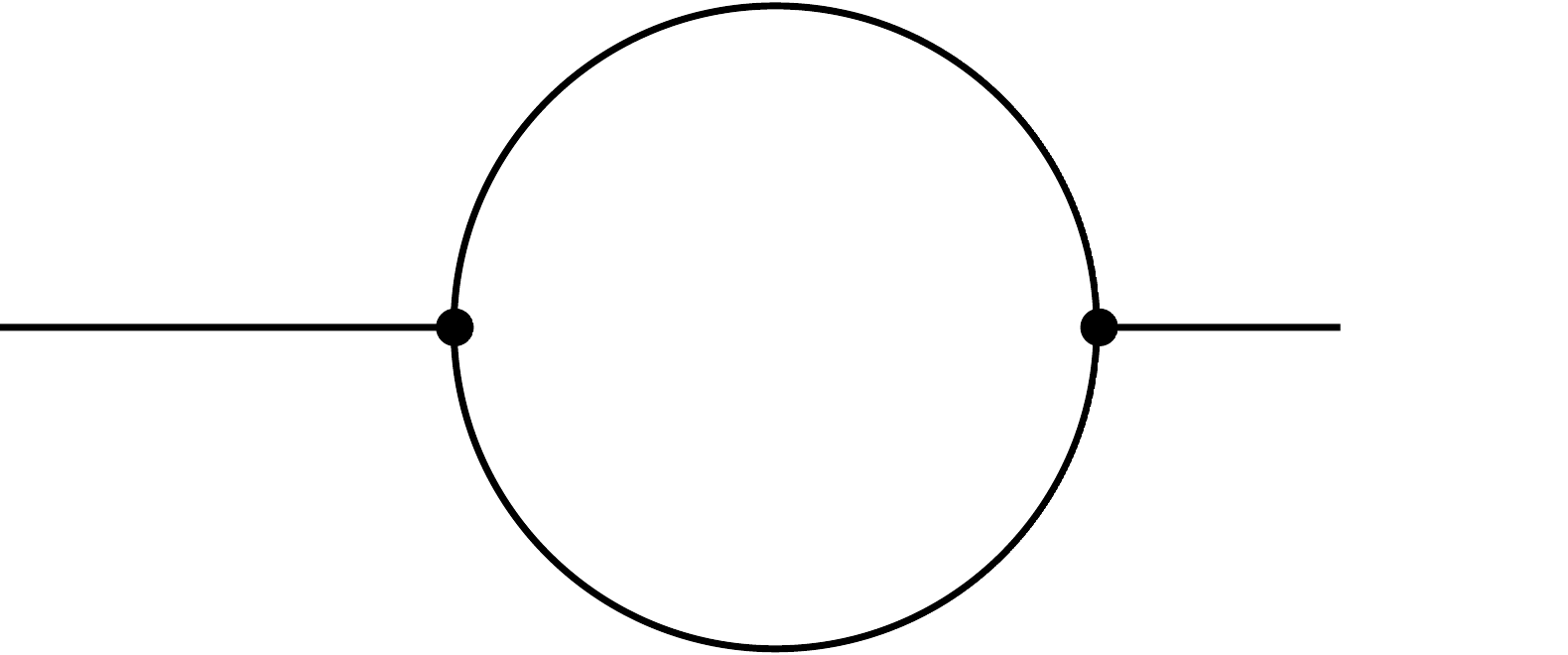}}
	\caption{Diagram ($I_{w1}$) contributing to wave function renormalization at one loop.}
	\label{Wawe1Renormalization}
\end{figure}

To obtain wave function renormalization we will consider renormalization of two-point 1PI Green function $\Gamma_2 (p^2)$, in fact of its derivative at $p^2=0$. At one loop there is only one diagram contributing, see Fig.\ref{Wawe1Renormalization}.
\begin{equation} 
\frac{\partial}{\partial p^2}I_{w1}\Bigg|_{p^2=0}=-\frac{\lambda^2\hat{\varphi}^2(4 \pi e^{-\gamma_E})^{\ep}}{12\hat{m}^2(4\pi)^4}
\end{equation}
This expression is free of singularities, and in Coleman-Weinberg scheme  \eqref{renormCWconditions} we get finite wave function renormalization at one-loop 
\begin{equation} 
\delta Z_1^{(CW)}=-\frac{\lambda(e^{-\gamma_E} \pi \gamma^{-2})^{\ep}}{64\pi^2}
\end{equation}
while in $\MSbar$ scheme this constant is zero:
\begin{equation} 
\delta Z_1^{\MSbar}=0
\end{equation}

\begin{figure}[h]
	\center{\includegraphics[width=0.67\textwidth]{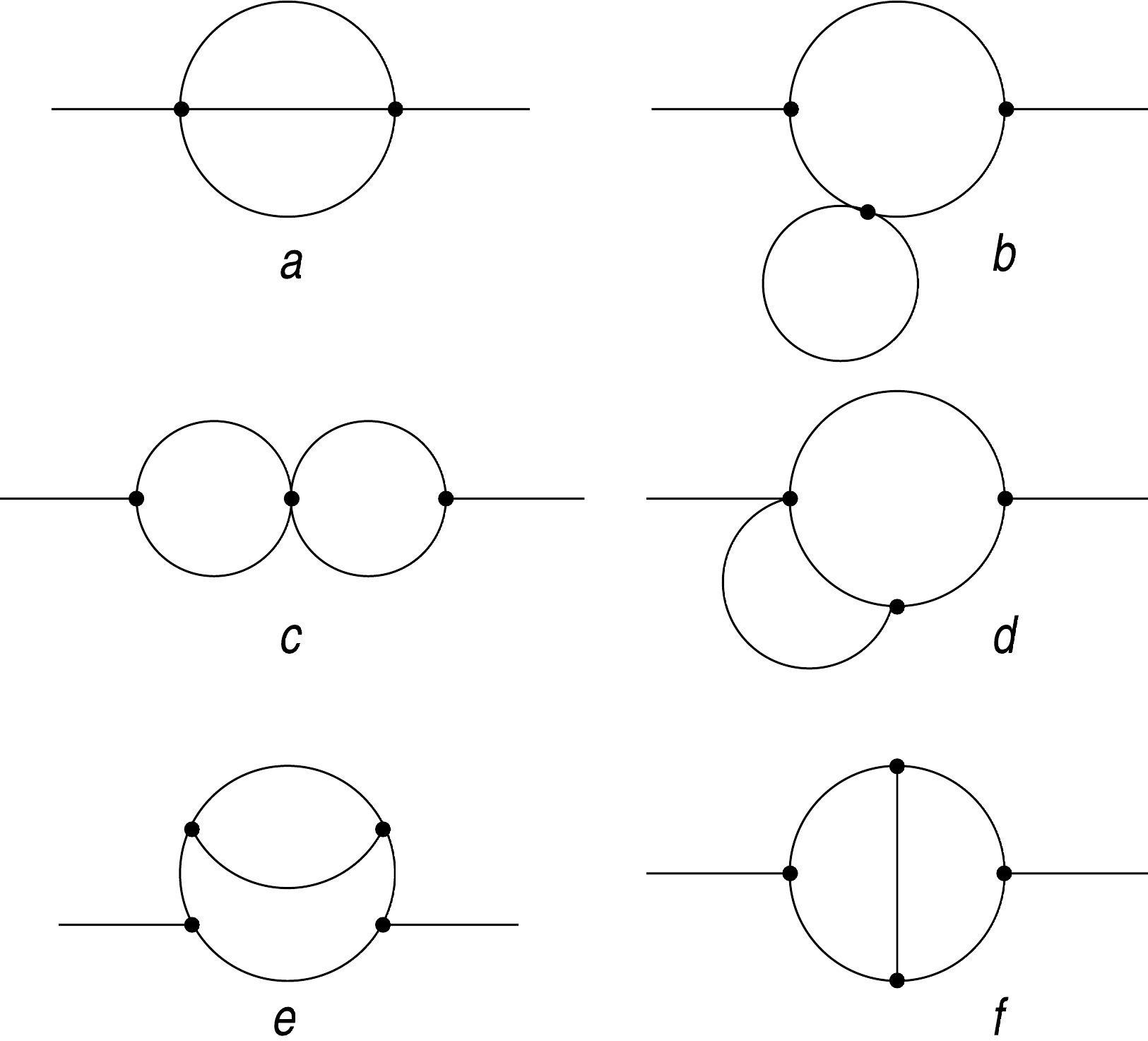}}
	\caption{Diagrams contributing to wave function renormalization at two loops.}
	\label{Wawe2Renormalization}
\end{figure}

At two loops there are six diagrams already, see Fig.\ref{Wawe2Renormalization}. Fortunately, we do not need to calculate these diagrams as a functions of $p^2$, all what is needed is to calculate their derivatives with respect to $p^2$ at $p^2=0$. In this case the calculation is easy and reduces as in the case of effective potential to 2-loop single-scale tadpole diagrams. As a result we have
\begin{equation} 
\frac{\partial}{\partial p^2}I_a\Bigg|_{p^2=0}=-\frac{\lambda^2(4 \pi e^{-\gamma_E})^{2\ep}}{(4\pi)^4}\left(\frac{1}{24\ep}+\frac{1}{16}-\frac{S_2}{2}\right)
\end{equation}
\begin{equation} 
\frac{\partial}{\partial p^2}I_b\Bigg|_{p^2=0}=-\frac{\lambda^3\hat{\varphi}^2(4 \pi e^{-\gamma_E})^{2\ep}}{12\hat{m}^2(4\pi)^4}\left(\frac{1}{2\ep}+1\right)
\end{equation}
\begin{equation} 
\frac{\partial}{\partial p^2}I_c\Bigg|_{p^2=0}=\frac{\lambda^3\hat{\varphi}^2(4 \pi e^{-\gamma_E})^{2\ep}}{12\hat{m}^2(4\pi)^4\ep}
\end{equation}
\begin{equation} 
\frac{\partial}{\partial p^2}I_d\Bigg|_{p^2=0}=\frac{\lambda^3\hat{\varphi}^2(4 \pi e^{-\gamma_E})^{2\ep}}{\hat{m}^2(4\pi)^4}\left(\frac{1}{6\ep}+\frac{2}{9}-S_2\right)
\end{equation}
\begin{equation} 
\frac{\partial}{\partial p^2}I_e\Bigg|_{p^2=0}=-\frac{\lambda^4\hat{\varphi}^4(4 \pi e^{-\gamma_E})^{2\ep}}{\hat{m}^4(4\pi)^4}\left(\frac{1}{24\ep}+\frac{53}{432}-\frac{S_2}{3}\right)
\end{equation}
\begin{equation} 
\frac{\partial}{\partial p^2}I_f\Bigg|_{p^2=0}=\frac{\lambda^4\hat{\varphi}^4(4 \pi e^{-\gamma_E})^{2\ep}}{\hat{m}^4(4\pi)^4}\left(\frac{2}{27}-\frac{7 S_2}{12}\right) ,
\end{equation}
where $S_2=\frac{4}{9\sqrt{3}} \Im\left(\text{Li}_2\left(e^{\frac{i \pi }{3}}\right)\right)$
Summing all diagrams above plus diagrams with one-loop counterterm insertions not shown in  Fig.\ref{Wawe2Renormalization} and using CW renormalization scheme prescription  
\eqref{renormCWconditions} for 2-loop wave function renormalization counterterm we get
\begin{equation} 
\delta Z^{(2)}_{CW}=-\frac{(e^{-\gE}\pi\gamma^{-2})^{2\ep}\lambda^2}{3(8\pi)^4}\left(\frac{2}{\ep}-137+\frac{304 \Im\left(\text{Li}_2\left(e^{\frac{i \pi }{3}}\right)\right)}{3\sqrt{3}}\right)
\end{equation}
Similarly in the case of $\MSbar$ scheme we have
\begin{equation} 
\delta Z_2^{\MSbar}=-\frac{\lambda^2(4 \pi e^{-\gamma_E})^{2\ep}}{24(4\pi)^4\ep} ,
\end{equation}
which is again in agreement with previously obtained results \cite{FunctionalMethodsPerturbationTheory,ScalingBehaviorPhi4,AnApproachEffectivePotential}.

\section{Sunset diagram}\label{sunset-appendix}

The most complicated diagram among 2-loop diagrams contributing to false vacuum decay is the sunset diagram $a11$ in Fig. \ref{FD}. The expression for the latter is given by 
\begin{multline} 
I_{sun}=\frac{(2\gamma\sqrt{3\lambda})^2}{12}\int d z_{\bot}\int d z'_{\bot}\int d^{(d-1)}\mathbf{z_{\|}}\int d^{(d-1)}\mathbf{z'_{\|}}\int\frac{d^{d-1}\mathbf{k_1}}{(2\pi)^{d-1}}\int\frac{d^{d-1}\mathbf{k_2}}{(2\pi)^{d-1}}\int\frac{d^{d-1}\mathbf{k_3}}{(2\pi)^{d-1}}\times
\nonumber \\
\times\Big\{G(z_{\bot},z'_{\bot},\mathbf{k_1})G(z_{\bot},z'_{\bot},\mathbf{k_2})G(z_{\bot},z'_{\bot},\mathbf{k_3})\tanh(\gamma z_{\bot})\tanh(\gamma z'_{\bot}) \nonumber \\
- G_{FV}(z_{\bot},z'_{\bot},\mathbf{k_1})G_{FV}(z_{\bot},z'_{\bot},\mathbf{k_2})G_{FV}(z_{\bot},z'_{\bot},\mathbf{k_3}) \Big\}e^{i\mathbf{(k_1+k_2+k_3) (z_{\|}-z'_{\|})}}
\end{multline}
Performing change of variables $x = \tanh (\gamma z_{\bot})$, $y = \tanh (\gamma z'_{\bot})$ and integrating over coordinates parallel to the bubble surface $\mathbf{z_{\|}}$ and $\mathbf{z'_{\|}}$ we get 
\begin{multline}
I_{sun} = -\frac{\lambda^2}{2} S_b \gamma^{-4\ep} \int_{-1}^1\int_{-1}^1 \frac{dx dy}{(1-x^2)(1-y^2)}\int\frac{d^{d-1}\bk_1}{(2\pi)^{d-1}}\int\frac{d^{d-1}\bk_2}{(2\pi)^{d-1}} \\ \times
\Big\{ 
x y G(x, y, \bk_1) G(x, y, \bk_2) G(x, y, \bk_3) - G_{FV}(x, y, \bk_1) G_{FV}(x, y, \bk_2) G_{FV}(x, y, \bk_3) 
\Big\} \\
\end{multline}
with $\bk_3 = \bk_1 + \bk_2$. To evaluate this integral it is convenient to modify Green function $G(x,y,\bk)$ as 
\begin{multline}
G (x, y, \bk) = \, \frac{1}{2\gamma m} \Bigg\{ \theta (x-y) \left(
\frac{1-x}{1+x}
\right)^{\frac{m}{2}} \left(
\frac{1+y}{1-y}
\right)^{\frac{m}{2}} \left(
1 - 3 \frac{(1-x)(1+m+x)}{(\frac{\gamma_m}{\gamma_M}+m)(\frac{2\gamma_m}{\gamma_M}+m)}
\right) \nonumber \\
\times \left(
1 - 3 \frac{(1-y)(1-m+y)}{(\frac{\gamma_m}{\gamma_M}-m)(\frac{2\gamma_m}{\gamma_M}-m)}
\right) + (x\to y)
\Bigg\}\, , 
\end{multline}
where $m = \frac{1}{\gamma_M}\sqrt{\bk^2 + 4\gamma_M^2}$. The original integral is recovered in the limit $\gamma_m = \gamma_m = \gamma$. Still, for $\gamma_m < \gamma_M$ we may evaluate the modified integral as a series in a ratio $\frac{\gamma_m}{\gamma_M}$ and provided it is convergent continue it to the point $\frac{\gamma_m}{\gamma_M} = 1$. It turns out that it is indeed the case. Moreover to obtained desired expansion we may use a strategy of regions, see \cite{StrategyRegions1,StrategyRegions2,StrategyRegions3,StrategyRegions4} and references therein. In our particular case only one region contributes, namely the one with $\bk_1 \sim \bk_2\sim \gamma_M$, where we may do the usual Taylor expansion of the integrand in $\gamma_m$. Our integrand contains the factors 
\begin{align}
\left(\frac{1-x}{1+x}\right)^{M/2} \left(\frac{1+y}{1-y}\right)^{M/2}\quad
\text{or} \quad \left(\frac{1-y}{1+y}\right)^{M/2} \left(\frac{1+x}{1-x}\right)^{M/2}\, ,
\end{align}
where $M = m_1 + m_2 + m_2$ and $m_i = \frac{1}{\gamma_M}\sqrt{\bk_i^2 + 4\gamma_M^2}$ and in the mentioned region for large $\gamma_M$ $M$ is also could be considered large. So, to evaluate integrals over $x$, $y$ variables we may use an expansion at $M\to \infty$ and saddle point approximation.  This way we set
\begin{align}
y = x + \frac{z}{M}
\end{align}
and Taylor expand integrands at $z=0$. Now, taking into account that for $M\to\infty$:
\begin{equation}
\begin{cases}
z \in (-\infty , 0) & x > y\\
z \in (0, \infty) & y > x
\end{cases}
\end{equation}
and integrating over $z$ variable we will for example obtain
\begin{align}
&\int_{-1}^1 \frac{d y}{1-y^2} \theta (x-y) \left(\frac{1-x}{1+x}\right)^{M/2} \left(\frac{1+y}{1-y}\right)^{M/2} = \frac{1}{M}\, , \\
&\int_{-1}^1 \frac{d y}{1-y^2} \theta (y-x) \left(\frac{1-y}{1+y}\right)^{M/2} \left(\frac{1+x}{1-x}\right)^{M/2} = \frac{1}{M} \, .
\end{align}
Finally taking integrals over $x$ and $y$ we get
\begin{multline}
I_{sun} = \frac{\lambda^2}{4}S_b \gamma^{-4\ep} \frac{1}{(2\pi)^{2 d-2}}\int d^{d-1}\bk_1 \int d^{d-1}\bk_2 \frac{1}{M m_1 m_2 m_3} \\
\times \left\{
1 - \frac{1}{m_1^2} - \frac{1}{m_2^2} - \frac{1}{m_3^2} - \frac{1}{m_1 M} -\frac{1}{m_2 M} - \frac{1}{m_3 M} + \frac{2}{3 M^2} + \ldots 
\right\} \label{sunset-series}
\end{multline}
Here $\dots$ denote higher order terms in the expansion. To further evaluate integrals over $\bk_1$ and $\bk_2$ we used Mellin-Barnes representation for the integral
\begin{equation}
\int d^d k_1 \int d^d k_2 \frac{1}{m_1^{a_1} m_2^{a_2} m_3^{a_3} M^{a_4}}\, ,
\end{equation}
where $d=3-2\ep$, $m_i = (k_i^2+4)^{1/2}$, $M=m_1+m_2+m_3$ as before and $a_i$ are arbitrary indexes. The latter could be easily derived and is given by 
\begin{multline}
\int d^d k_1 \int d^d k_2 \frac{1}{m_1^{a_1} m_2^{a_2} m_3^{a_3} M^{a_4}} = \frac{\pi^d 2^{2 d-a_1-a_2-a_3-a_4}}{(2\pi i)^3\Gamma (a_4)\Gamma\left(\frac{d}{2}\right)}\int_{-i\infty}^{i\infty} dz_1 \int_{-i\infty}^{i\infty} dz_2 \int_{-i\infty}^{i\infty} dz_3  \\
\times \frac{\Gamma (-z_1) \Gamma (-z_2) \Gamma (-z_3) \Gamma (a_4 + z_1+z_2)\Gamma\left(\frac{a_3+a_4+z_1+z_2+2z_3}{2}\right)}{\Gamma \left(
\frac{a_3+a_4+z_1+z_2}{2}	
\right)\Gamma\left(\frac{a_1-z_1}{2}\right)\Gamma\left(\frac{a_2-z_2}{2}\right)} \\
\times \frac{\Gamma \left(\frac{a_1+a_3+a_4+z_2+2 z_3-d}{2}\right)\Gamma \left(\frac{a_2+a_3+a_4+z_1+2 z_3-d}{2}\right)\Gamma\left(
\frac{d-a_3-a_4-z_1-z_2-2z_3}{2}	
\right)\Gamma\left(\frac{a_1+a_2+a_3+a_4+2z_3-2d}{2}\right)}{\Gamma\left(
\frac{a_1+a_2+z_1+z_2}{2} + a_3 + a_4 + 2z_3 -d
\right)}\, .
\end{multline}
The required Mellin-Barnes integrals with specific values of propagator indexes are then  evaluated numerically with the help of \cite{Czakon}. Finally for the sunset diagram we obtained
\begin{equation} 
\label{sunset_final}
I_{sun}=\frac{S_c\lambda^2(e^{-\gamma_E} \pi \gamma^{-2})^{2\ep}}{8\pi^4}\left[\frac{9}{64\ep^2}+\frac{s_{-1}}{\ep}+s_0\right]\, ,
\end{equation}
where $s_{-1} = 0.39522$ and $s_0\approx 0.71$. The value of $s_{-1} = 0.39522$ constant which we got from the first $20$ terms of the series\footnote{We have written explicitly only two of them in \eqref{sunset-series}} \eqref{sunset-series} is actually several percent less then the exact value which could be found from the cancellation of $1/\ep$ poles in the process of renormalization 
\begin{equation}
s_{-1} = \frac{197}{384} - \frac{3\sqrt{3}\pi}{160}\approx 0.410995
\end{equation}
which is due to slow convergences of the mentioned series. It could be certainly further improved but this goes beyond the goal of this paper.

\bibliographystyle{ieeetr}
\bibliography{litr}

\end{document}